\documentclass{JHEP3}
\let\ifpdf\relax
\usepackage{ifpdf}
\pdfoutput=1
\usepackage{graphicx}
\usepackage{fancyvrb}
\usepackage{makeidx}
\usepackage{latexsym,amssymb,amsmath}
\usepackage{booktabs}

\newcommand{\be}{\begin{equation}}
\DefineVerbatimEnvironment{code}{Verbatim}{fontsize=\small}
\newcommand{\ee}{\end{equation}}
\newcommand{\bi}{\begin{itemize}}
\newcommand{\ei}{\end{itemize}}
\makeindex 
\newcommand{\met}{\mbox{${\rm \not\! E}_{\rm T}$}}

\newcommand{\Sec}[1]{Sec.~\ref{#1}}
\newcommand{\Fig}[1]{Fig.~\ref{#1}}

\newcommand{\Ref}[1]{Ref.~\cite{#1}}

\def\gev{~{\mbox{GeV}}}

\title{(Light) Stop Signs}
\author{Zhenyu Han, Andrey Katz, David Krohn, and Matthew Reece\\
 Department of Physics, Harvard University, Cambridge, MA 02138\\
 E-mail: \email{zhan,andrey,dkrohn,mreece@physics.harvard.edu}}
\abstract{Stop squarks with a mass just above the top's and which decay to a nearly massless LSP are difficult to probe because of the large SM di-top background.
Here we discuss search strategies which could be used to set more stringent bounds in this difficult region.
In particular, we note that both the rapidity difference $\Delta y(t,{\bar t})$ and spin correlations (inferred from, for example, $\Delta \phi(\ell^+,\ell^-)$) are sensitive to the presence of stops. We emphasize that systematic uncertainties in top quark production can confound analyses looking for stops, making theoretical and experimental progress on the understanding of Standard Model top production at high precision a very important task. We estimate that spin correlation alone, which is relatively robust against such systematic uncertainties, can exclude a 200 GeV stop at 95\% confidence with 20 fb$^{-1}$ at the 8 TeV LHC.}
\begin{document}
%\maketitle
%\tableofcontents

%%%%%%%%%%%%%%%%%%%%%%%%%%%%%%%%%%%%%%%%%%%%%%%%%%%%%%%%%
%\section{Stop Sneakin' Round: Introduction}
\section{Introduction}
%%%%%%%%%%%%%%%%%%%%%%%%%%%%%%%%%%%%%%%%%%%%%%%%%%%%%%%%%

Supersymmetry remains the most compelling theoretical explanation for how the vast hierarchy between weak and Planck scales can persist in the face of quantum mechanical effects. The leading quantum corrections that make the Standard Model an unnatural theory are the quadratically divergent contributions of top quark loops to the Higgs boson mass term, which are largely canceled in supersymmetric theories by loops of scalar tops. Minimizing fine tuning in supersymmetric theories, then, requires that the stop mass be as close to the top mass as possible. From the theoretical viewpoint, this motivates consideration of models in which the stop may be among the lightest superpartners~\cite{Dimopoulos:1995mi,Cohen:1996vb}. 

The LHC is setting ever more stringent bounds on SUSY.  Very roughly, if the first and second generation squarks are all of the same mass and they decay as $\tilde{q}\rightarrow q+\chi^0$ the
current bounds require $m_{\tilde{q}}\gtrsim 1.3~{\rm TeV}$ (for decoupled gluinos), and if the gluinos decay as $\tilde{g}\rightarrow q\bar q+\chi^0$ the bound is roughly $m_{\tilde{g}}\gtrsim 1~{\rm TeV}$ (for decoupled squarks).  These bounds come primarily from searches for jets + MET~\cite{Aad:2011ib,CMS-PAS-SUS-11-004,ATLAS-CONF-2012-033,CMS-PAS-SUS-12-011}, and become stronger if squarks and gluinos have comparable masses. These results force ``vanilla" supersymmetry, meaning the class of well-studied models with flavor universal squark masses and large amounts of missing transverse momentum in decays, into an uncomfortably unnatural corner. 

However, current data allows another interesting possibility. If squarks of the third generation are significantly lighter than those of the first two, the bounds on the lighter particles can be considerably weaker~\cite{Essig:2011qg,Kats:2011qh,Brust:2011tb,Papucci:2011wy,Bi:2011ha,Desai:2011th}. 
Sbottoms which decay as $\tilde{b}\rightarrow b+\chi^0$, for instance,
are only constrained to  $m_{\tilde{b}}\gtrsim 400~{\rm GeV}$ for light LSPs, with the bounds becoming much less stringent as $m_{\chi^0}$ is increased~\cite{Aad:2011cw}.
Perhaps the weakest bounds are on the stop, which after $1~{\rm fb}^{-1}$ is still allowed to be at or perhaps slightly below the top mass, according to theoretical estimates~\cite{Kats:2011it,Kats:2011qh} that carry too much uncertainty in detector modeling to make a sharp claim. Third generation squarks produced indirectly in gluino decays lead to signatures of same-sign dileptons or of jets and MET, constraining the gluino to be heavier than about $850~{\rm to}~950$~GeV~\cite{ATLAS-CONF-2012-004,Collaboration:2012sa,CMS-PAS-SUS-11-027,ATLAS-CONF-2012-037}, which is still compatible with naturalness.\footnote{For theoretical models which motivate a closer look at SUSY with light third generation superpartners see e.g.~\cite{Sundrum:2009gv,Barbieri:2010pd,Redi:2010yv,Craig:2011yk,Csaki:2011xn,Csaki:2012fh,Craig:2012di}.}

These considerations strongly motivate a deeper study of the phenomenology of light stops. The study of simplified spectra, based on naturalness considerations, in which a stop decays to a top and neutralino was advocated in~\cite{Meade:2006dw}, and a number of related works have appeared since~\cite{Han:2008gy,Plehn:2010st,Plehn:2011tf,Bai:2012gs}, as well as variations in which the stop plays a role in coannihilation to explain dark matter abundance~\cite{Bhattacharyya:2011se,Choudhury:2012tc} or has flavor-violating couplings to explain top $A_{FB}$~\cite{Isidori:2011dp}. (We have recently learned of two other papers in preparation~\cite{Alves:new,Kaplan:new} that take different approaches to the problem, which may be complementary to ours.) The challenge of stop phenomenology is to disentangle the stop signal from top backgrounds. For our purposes, it is useful to introduce a division of ${\tilde t} \to b W^+ \chi^0$ signals into three categories depending on the stop mass:
\begin{itemize}
\item {\em Three-body stops}: if stops are light enough, they cannot decay to an on-shell top quark and a neutralino (or gravitino), so the three-body decay ${\tilde t} \to bW^+\chi^0$ dominates. (With more extreme squeezing of the spectrum, a four-body decay or flavor-violating decay may result, but we will not consider this limit.) Many kinematic distributions such as $m_{\ell b}$ or transverse mass for stop decay will be significantly different from those in top decay~\cite{Chou:1999zb,Kats:2011it}. However, this region also suffers from relatively low acceptance, and probably deserves a closer look in the future. 
\item {\em Two-body stops}: if $m_{\tilde t} \gg m_t$, the decay is two-body, ${\tilde t} \to t\chi^0$, and the invisible neutralino or gravitino can carry away a large amount of momentum. The simplest way to exploit this is to impose a hard MET cut. More sophisticated approaches use the fact that MET in the top background arises from neutrinos in $W$ decay, which suggests the power of a transverse mass cut in the semileptonic case~\cite{Han:2008gy} and a cut on $M_{T2}$ computed with the two leptons and MET in the dileptonic case~\cite{Cohen:2010wv,Kats:2011qh}. A recent update on this regime appeared in Ref.~\cite{Plehn:2012pr}.
\item {\em Stealth (``one-body") stops}: if the neutralino (or gravitino) is light and $m_{\tilde t}$ is only slightly larger than $m_t$, two-body decays ${\tilde t} \to t\chi^0$ in which the momentum of $\chi^0$ is very small can predominate. Kinematically, these are effectively ``one-body decays" since the top carries all the momentum, and stop pair production can be extremely difficult to separate from top pair production~\cite{Fan:2011yu}. Furthermore, unlike compressed supersymmetry scenarios, the events do not become more distinctive when recoiling against an additional hard jet~\cite{Fan:2012jf}.
\end{itemize}
The stealth stop regime is the most challenging and can involve a large new physics cross section at the LHC. This regime is the focus of our current study. 

\FIGURE[ht]{
\includegraphics[width = 0.6\textwidth]{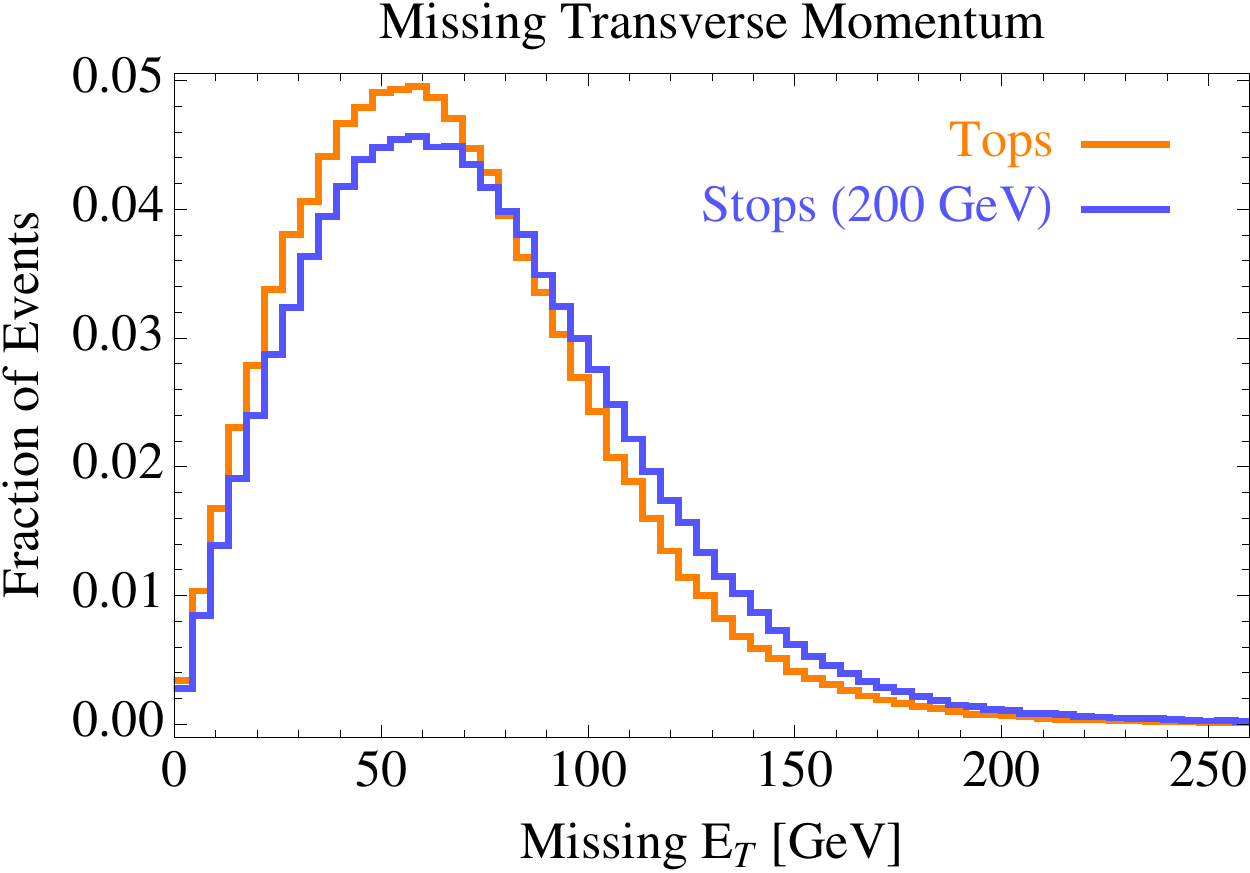}
\caption{$\met$ distribution in top and stop events, where we have considered stop decays to massless neutralinos. The rate is normalized to the number of events with two isolated leptons.    }
\label{fig:metnorman}
}

We illustrate the stealth regime in Fig.~\ref{fig:metnorman}, which shows the missing transverse energy distribution for dileptonic events from top pairs and 200 GeV stop pairs (decaying as ${\tilde t} \to t \chi^0$). This is based on a simulation with cuts that we will describe in Sec.~\ref{subsec:jetlevel}. The distributions for tops and stops are very similar, because in the rest frame of the stop, in the limit of small mass difference and massless $\chi^0$, the momentum of the decay products is $\approx \delta m = m_{\tilde t} - m_t$. In the lab frame, the $\chi^0$ carries away invisible momentum of order $\gamma~\delta m$, and for production of typical stop pairs the boost is not large.

\FIGURE[ht]{
\includegraphics[scale=0.7]{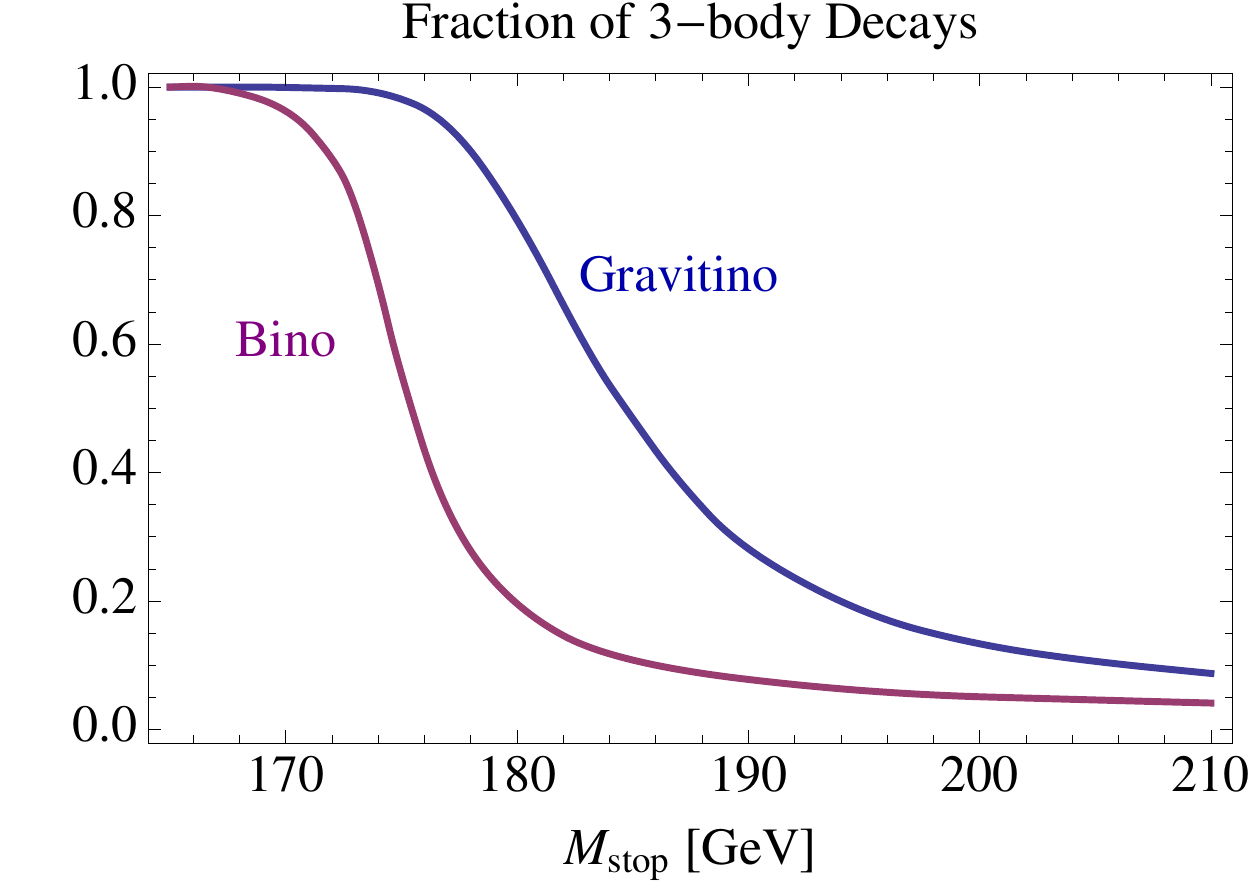}
\caption{Fraction of stop decays, ${\tilde t} \to W^+ b \chi$, which are three-body, as a function of the stop mass. More precisely, we are labeling a decay ``three-body" when $m(W^+ b) < m_t - 3 \Gamma_t$, and have taken the top quark mass to be 173 GeV. The neutral fermion $\chi$ is either the gravitino ${\tilde G}$ or a massless bino ${\tilde B}$. In the gravitino case, three-body decays persist for larger stop masses, so the ``maximally stealthy stop" is at masses nearer 200 GeV than 175 GeV.}
\label{fig:threebodyfrac}
}

If a stop decays to a massless neutralino, the transition from the three-body regime to the stealth regime is not smooth. The three-body decay ends abruptly at $m_{\tilde t} = m_t$, at which point two-body stealth decays dominate until the mass splitting becomes large enough that the decays are no longer stealthy. The case of a stop decay to a gravitino is slightly more subtle; the gravitino couples to SUSY breaking, leading to two extra powers of $m_{\tilde t} - m_t$ phase-space suppression in the two-body decay rate. This allows the three-body regime to extend to somewhat higher masses, as illustrated in Fig.~\ref{fig:threebodyfrac}. (This plot and others throughout the paper rely on simulations performed with MadGraph~5~\cite{Alwall:2011uj}, as well as goldstino vertices we have implemented~\cite{ReeceStopNLSPpage} using the UFO format~\cite{Degrande:2011ua}). The estimates in~\cite{Kats:2011qh} show that current analyses have weakened sensitivity in the range $m_t \lesssim m_{\tilde t} \lesssim 250~{\rm GeV}$, which we will take as our characterization of the stealth stop window. We review the current searches relevant for stops in Sec.~\ref{sec:current}, characterizing the extent to which they are simple top rate measurements in this window. Although  more data will reduce the statistical errors on measurements of the top, both experimental systematics and theoretical uncertainties will remain.  Measurements of the top are notoriously difficult (see, e.g., \Ref{Melnikov:2011qx}), and so the more handles one has to constrain/discover stops, the better.

\FIGURE[ht]{
\includegraphics[scale=0.7]{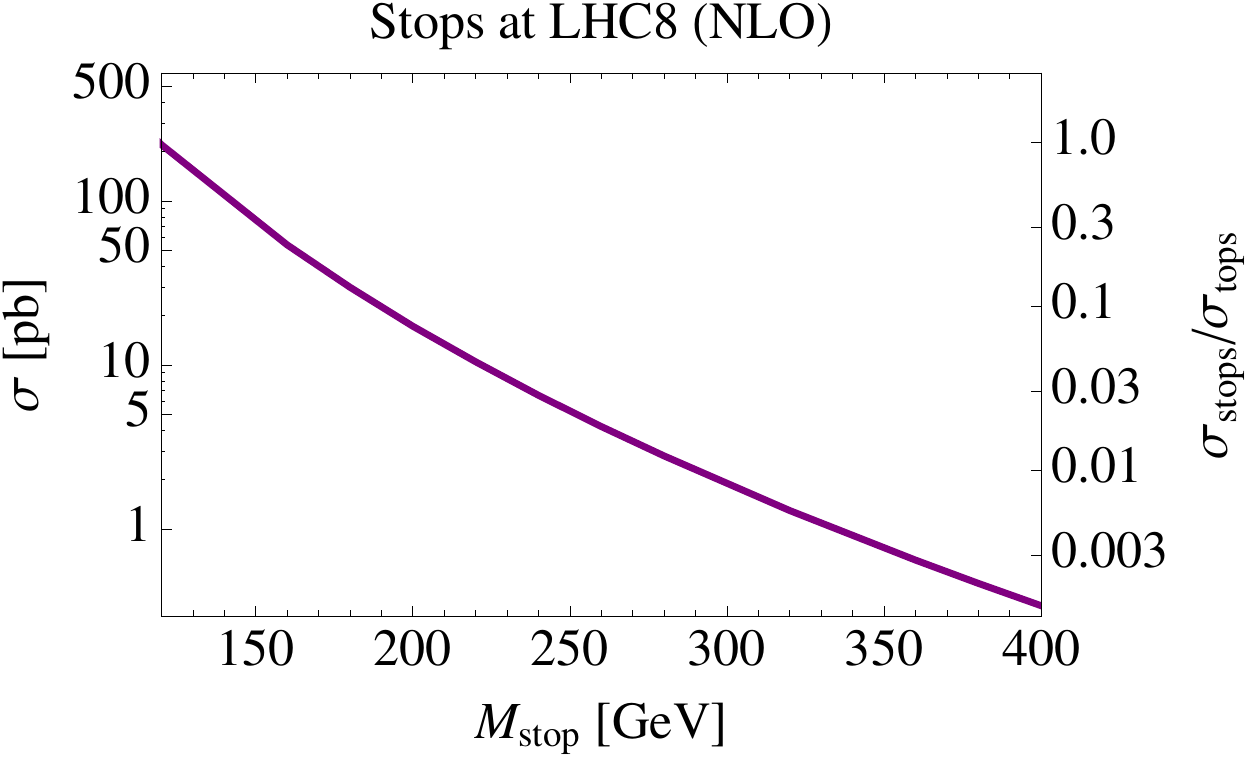}
\caption{NLO stop pair production cross section at the 8 TeV LHC, as reported by Prospino~\cite{Beenakker:1997ut}. The vertical axis on the right shows the rate as a ratio to the $t{\bar t}$ rate.}
\label{fig:stopcrosssection}
}

\begin{table}[ht]\footnotesize
\begin{center}
\begin{tabular} {@{}lp{11.5cm}@{}}
\toprule
{\bf Observable} & {\bf Feasibility} \\
\midrule
Top production rate & Stops near $m_t$ can increase  the measured value of $\sigma_{t\bar t}$ by up to $\sim 15\%$. However,  
(1) even at NNLL the uncertainties in $\sigma_{t\bar t}$  are  $\sim 10\%$ and (2) theoretical ambiguities 
in the interpretation of the measured value of $m_t$ can further increase the uncertainty by a similar degree.     \\ \addlinespace
\met-based quantities & While the LSP from stop decays contributes to the \met  in an event, this effect 
is usually greatly diluted by the contributions from neutrinos (again, for $m_{\tilde t}$ near $m_t$).   
When this is not the case ({\it e.g.},  with fully-hadronic top decays, or in the tails of $m_T$ distributions) 
we expect the experimental systematics to be very challenging.  \\ \addlinespace
% \met distributions therefore show very little diffence between tops and stops. \\  
$\Delta y(t,\bar t)$ & The shape of the $\Delta y(t,\bar t)$ distribution differs markedly  
between stop and top production.  However, once one accounts for the small stop production rate this effect can be mimicked 
by QCD NLO uncertainties. \\ \addlinespace
Spin correlations & Measurements of top spin correlation at the LHC are quite insensitive to higher order corrections, and show a clear difference between 
top and stop production.  While we expect this measurement to be statistically limited, we  conclude that this is 
one of the most robust channels to use in searching for light stops.
\\ 
\bottomrule
\end{tabular}
\end{center}
\caption{A summary of observables sensitive to the presence of stops above the SM di-top background, and comments 
on the feasibility of employing these in searching for light stops.  Note that when we speak of tops in stop events we are referring to stops reconstructed as  tops.}
\label{tab:feasibility}
\end{table}  

Here we present a set of search strategies which can be used to constrain stops in this difficult region of parameter space.  While we find no single smoking-gun signature for light stops, a few robust physical considerations can enhance the sensitivity of searches. It will be important to combine these considerations, because stops are rare. At the 7 TeV LHC, the top cross section is about 165 pb (increasing to about 230 pb with 8 TeV collisions~\cite{Cacciari:2011hy}), while the stop cross section for $m_{\tilde t} = m_t$ is only about one-sixth as large, and drops steeply at larger masses. (See Fig.~\ref{fig:stopcrosssection} for NLO results; recent, more accurate, calculations of stop production may be found in~\cite{Beenakker:2010nq}.) Thus, finding stops by simply measuring the total rate of events passing top selection cuts is a challenge. 

The most important physical handle on stops in the stealthy regime is the characteristic that allows them to play their divergence-canceling role in SUSY: they are bosons, not fermions. In particular, because they are spin zero, they are produced without spin correlations and the decay of a stop and antistop are completely uncorrelated. Top pair production, on the other hand, involves spin correlation effects, linking the decay angles on the two sides of the event. Thus, a top pair sample enriched by stealthy stops should have correlations that are washed out by an amount related to the stop cross section. The effect is small, but it still can be used, and we elaborate on these ideas in Sec.~\ref{sec:spin}. Even more important, it was shown in~\cite{Melnikov:2011ai} that full matrix element spin-correlation measurement is stable with respect to NLO corrections, suffering from very mild systematic uncertainties. Stability with respect to NLO corrections distinguishes this measurement from other methods.  

Other effects that we use to discriminate between these two samples have to do with smallness of $\tilde t {\bar {\tilde t}}$ production compared to $t\bar t$. Two effects conspire to suppress the production rate of stops compared to tops  near the threshold. First, the process $gg \to t \bar t$ has $t$- and $u$-channel singularities, which are regularized by $m_t$. There is no analogous singularity in $\tilde t {\tilde t}^\ast$ production; therefore $t \bar t$ production is enhanced, and $t \bar t$ events tend to have larger rapidity gap between tops, relative to ${\tilde t} {\tilde t}^\ast$. This is the most important stop rate suppression mechanism at the LHC. On the other hand, one can take advantage of the rapidity gap as a possible discriminator. We will elaborate on this potentially useful handle in section~\ref{sec:rapidity}, though we notice that this approach can suffer from unpleasantly large systematic uncertainties. Another mechanism which suppresses the stop production rate is a $p$-wave suppression in the $s$-channel stop production from a $q{\bar q}$ initial state, motivating a look at the higher $p_T$ regime, where the stop portion in the sample might be enhanced. This process is very important at the Tevatron, but less interesting at the LHC due to suppressed $q{\bar q}$ parton luminosities. We will further comment on this in Appendix~\ref{sec:boosted}. Although we do not show a single channel allowing for stealthy stop discovery, we claim that a combination of the above mentioned tools can give us more than 2$\sigma$ sensitivity to a stealthy stop by the end of $\sqrt{s} = 8 $ TeV run. We summarize the expected sensitivity in Table~\ref{tab:feasibility}. 

Our paper is organized as follows. In the next section we review the current searches for SUSY and further discuss the features of the stealth regime. In section~\ref{sec:spin} we discuss discrimination between tops and stops through spin correlation. In section~\ref{sec:rapidity} the rapidity gap and its practical use are discussed. (In Appendix~\ref{sec:multivariate} we briefly show how these tools can be combined to achieve better sensitivity.) Finally in Sec.~\ref{sec:discussion} we conclude. 

%%%%%%%%%%%%%%%%%%%%%%%%%%%%%%%%%%%%%%%%%%%%%%%%%%%%%%%%%
%\section{All the Old Showstoppers: Current LHC Searches}
\section{Current LHC Searches}
\label{sec:current}
%%%%%%%%%%%%%%%%%%%%%%%%%%%%%%%%%%%%%%%%%%%%%%%%%%%%%%%%%

Several existing LHC studies have some bearing on the question of light stops, and will become increasingly applicable with more data. In this section, we will survey them, arguing that currently the main discriminating variables being used are the top cross section, missing $E_T$, and more sophisticated proxies for missing $E_T$ (like transverse mass).

Because kinematics in the stealth regime resembles SM $t{\bar t}$ production, measurements of the top quark cross section potentially constrain stop production. One difficulty with attempting to find new physics simply by measuring the top cross section relative to its Standard Model value is that systematic uncertainties matter. As the top quark pole mass varies by about 5 GeV, the total cross section for $t{\bar t}$ production varies by about 15\%~\cite{Beneke:2011mq}. In order to make a convincing measurement of new physics, then, one needs both an accurate measurement of the total cross section and an accurate kinematic measurement of the top pole mass. Furthermore, the theoretical prediction of cross section given mass must be accurate enough to resolve a discrepancy. Current theoretical calculations of cross sections have 10\% errors from a combination of scale variation, $\alpha_s$, and PDF uncertainties~\cite{Cacciari:2011hy,Beneke:2011mq}. Measurements of the top mass from kinematics can be fraught with difficulties over the precise definition of ``top mass," with a common claim being that experiments have traditionally measured the ill-defined Pythia mass rather than the pole mass. Even the pole mass of the top is not well-defined, requiring a specification of the treatment of IR ambiguities due to confinement in QCD~\cite{Hoang:2008yj}. Inferences of the top quark pole mass from total cross section (using NNLO theory calculations~\cite{Langenfeld:2009wd}) at D$\varnothing$~\cite{Abazov:2011pta}, CMS~\cite{CMS-PAS-TOP-11-008}, and ATLAS~\cite{ATLAS-CONF-2011-054} yield $167.5^{+5.4}_{-4.9}$, $170.3^{+7.3}_{-6.7}$, and $166.4^{+7.8}_{-7.3}~{\rm GeV}$, while the latest Tevatron combination for the top mass inferred from kinematics is $173.2 \pm 0.9~{\rm GeV}$. A CMS cross section combination measured $165.8\pm2.2~{\rm(stat.)}\pm10.6~{\rm(syst.)}\pm7.8~{\rm(lumi.)}~{\rm pb}$~\cite{CMS-PAS-TOP-11-024}, whereas an ATLAS cross section combination measured $177\pm3~{\rm(stat.)}^{+8}_{-7}~{\rm(syst.)}\pm7~{\rm(lumi.)}$ pb~\cite{ATLAS-CONF-2012-024}. Taking all of this into account, it is clear that any apparent 15\% deviation in the top cross section would be much more plausibly attributed to uncertainties in $\alpha_s$, PDFs, $m_t$, luminosity, or any number of other systematic sources than to new physics. A claim of stop discovery {\em must} rely on kinematic and angular distributions, not simply rates.

\FIGURE[ht]{
\includegraphics[scale=0.7]{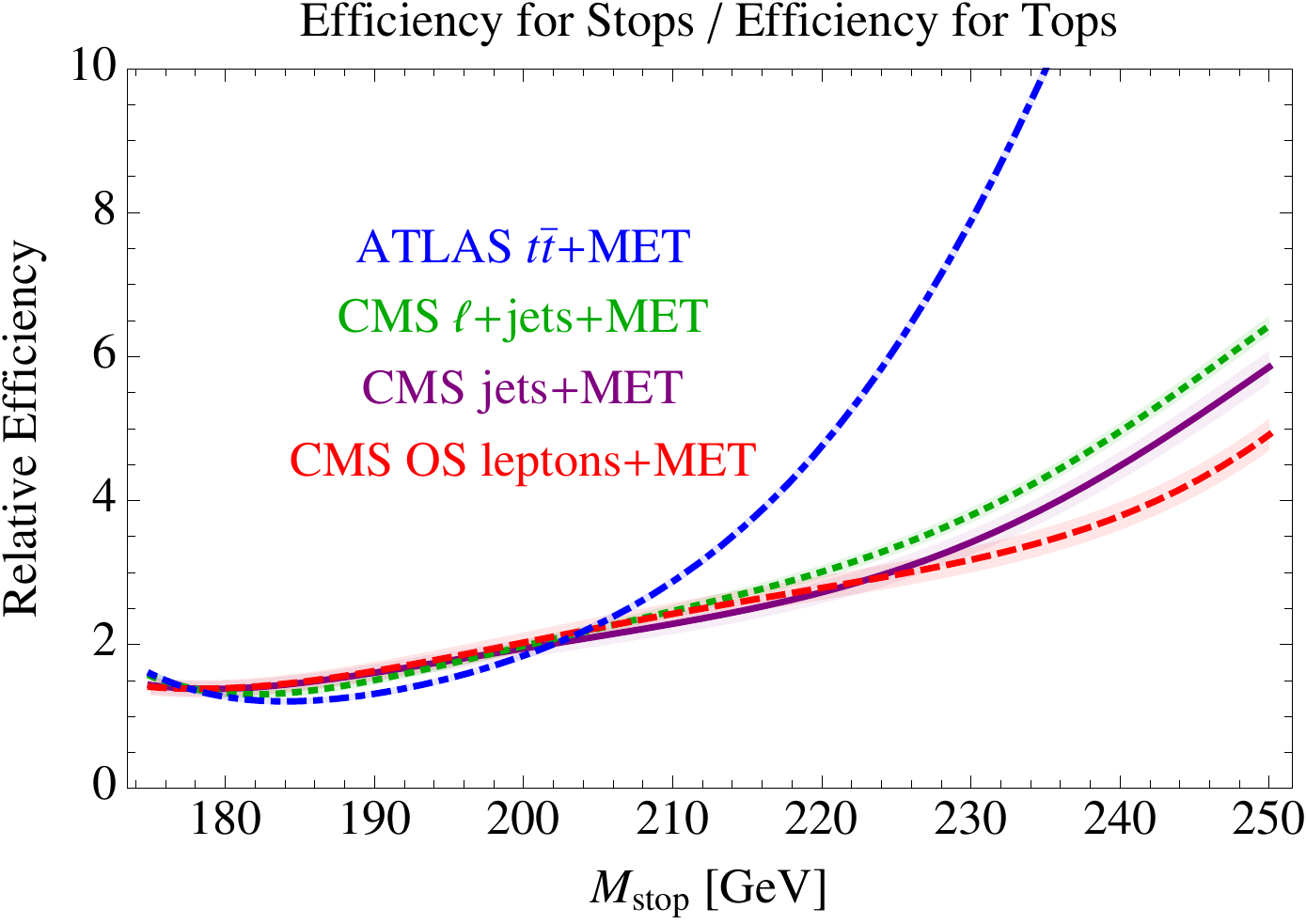}
\caption{Estimated relative efficiency of various existing studies for tops and for stops decaying to (on- or off-shell) top and massless bino. Notice that efficiencies for stops do not become significantly larger than for tops until the stop mass is above about 210 GeV; the regime below this is the especially challenging case of stealthy stops. At higher masses, stops are easier to separate from tops, although rates become small. (At 210 GeV, the Prospino estimate shown in Figure~\ref{fig:stopcrosssection} is $\sigma_{\rm stop} \approx 0.06 \sigma_{\rm top}$.)}
\label{fig:relativeefficiency}
}

Stops could also be constrained by various new physics searches, although these currently set weak limits, as discussed in~\cite{Kats:2011qh,Brust:2011tb,Papucci:2011wy}. We have computed the efficiency for stop pair production events to pass the cuts of some of the existing searches, and plotted its ratio to the efficiency for SM $t{\bar t}$ events in Figure~\ref{fig:relativeefficiency} as a function of the stop mass. (The simulation makes use of Pythia~\cite{Sjostrand:2006za,Sjostrand:2007gs} and FastJet~\cite{Cacciari:2005hq,Cacciari:2011ma}.) As representative examples, we have plotted the medium cuts of the CMS all-hadronic search~\cite{CMS-PAS-SUS-11-004}, the $250-350~S^{\rm lep}_T$ bin of the CMS lepton projection analysis~\cite{CMS-PAS-SUS-11-015}, the high-MET selection of the CMS opposite-sign dileptons analysis~\cite{CMS-PAS-SUS-11-011}, and the ATLAS $t{\bar t}+{\rm MET}$ analysis~\cite{Aad:2011wc}. The estimate of the reach from these searches presented in~\cite{Kats:2011qh} showed a possible limit near the top mass from the CMS $\ell+{\rm jets}+{\rm MET}$ search~\cite{CMS-PAS-SUS-11-015}, and a hope for future exclusions above 250 GeV from the ATLAS $t{\bar t}+{\rm MET}$ search~\cite{Aad:2011wc}. Moreover, the estimates of~\cite{Brust:2011tb} showed a borderline exclusion of a single stop with a mass around 300 GeV by the ATLAS search~\cite{Aad:2011wc}. The discrepancies in estimates between Refs.~\cite{Kats:2011qh} and~\cite{Brust:2011tb} are very minor and can possibly be explained by details of the simulation and finally resolved by an updated experimental search.     

There are two interesting features to note in Figure~\ref{fig:relativeefficiency}. The first is that, for stops heavier than about 240 GeV, it is straightforward to build a search that has much higher efficiency for identifying stop events than top events. In particular, the ATLAS $t{\bar t}+{\rm MET}$ search does very well by focusing on the semileptonic channel and requiring a transverse mass cut $m_T(\ell, \met) > 150$ GeV. For semileptonic top events, one expects this transverse mass variable to be dominated by the neutrino from a $W$ decay and typically bounded above by 80 GeV. A significant background from dileptonic top events where one lepton is lost remains, but this too can be rejected by a clever variation on $M_{T2}$~\cite{Bai:2012gs}. Similarly, one could search in the dileptonic channel using $M_{T2}$ constructed with two leptons and missing $E_T$ to reject $t{\bar t}$ backgrounds, again because it has an edge at the $W$ mass~\cite{Cohen:2010wv, Kats:2011qh}. Thus, the lesson to take away from the dramatic rise in the blue dot-dashed curve in Figure~\ref{fig:relativeefficiency} is that cutting on kinematic variables that are bounded above in the background can be very effective, and generalizations of these variables will be an interesting tool to apply in the future.

\FIGURE[ht]{
\includegraphics[scale=0.7]{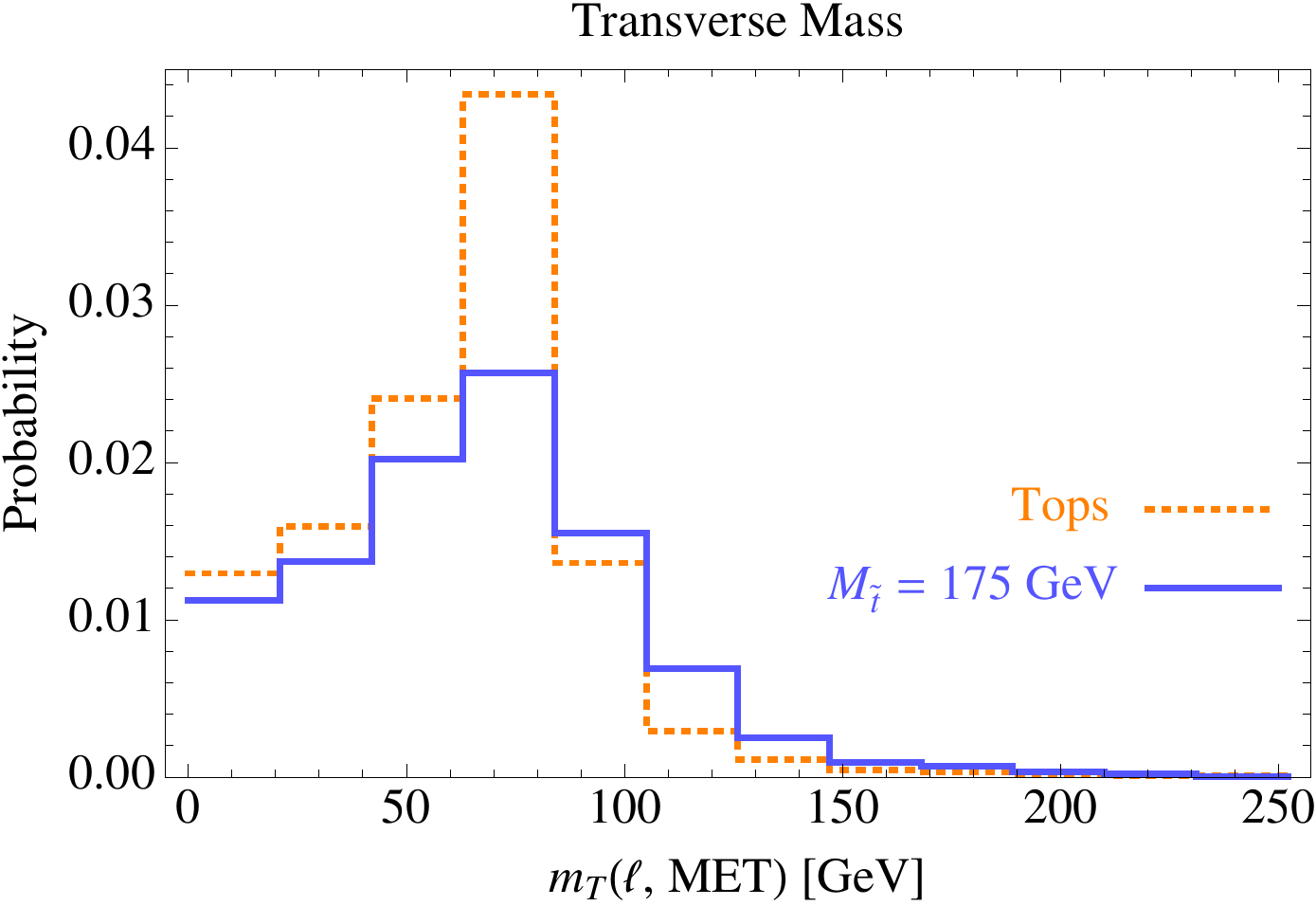}
\caption{Fraction of events with transverse mass $m_T$ in a given range and passing jet and lepton selection criteria from the ATLAS $t{\bar t}+{\rm MET}$ study~\cite{Aad:2011wc}. Notice that stop events have a larger tail above $M_W$ than top events do, even though the stop mass is barely above the top mass. This is due to three body decays ${\tilde t} \to bW^+{\tilde \chi}^0_1$. In the large $m_T$ bins, the efficiency for stops is about twice the efficiency of tops.}
\label{fig:transversemass}
}

The second feature of Figure~\ref{fig:relativeefficiency} that is interesting is the stealth regime, roughly $175~{\rm GeV} < M_{\rm stop} < 220~{\rm GeV}$ (the upper limit of this regime is somewhat subjective). One might have expected that in this regime, the efficiency for stops would be almost identical to that for tops. We see in the figure that this isn't true, even for $M_{\rm stop} < 180$ GeV; the efficiency of the selection cuts for stops very near the top mass is about 30\% to 60\% higher than that for tops, depending on the precise stop mass and the study considered. The reason is that, in the stealth regime, the bulk of the kinematic distributions for tops and stops are similar but the tails can still be significantly different. All of the new physics searches we have plotted involve large missing $E_T$ requirements, which are more easily satisfied in the stop events than top events. In particular, for stops barely above the top mass, we have checked that the enhancement of the stop efficiency with the ATLAS $t{\bar t}+{\rm MET}$ selection cuts is explained by events in which at least one stop decay is 3-body, with $m(bW) \ll m_t$, allowing the bino to carry away significantly more missing momentum than stealth kinematics would predict. The resulting difference in transverse mass distributions is shown in Figure~\ref{fig:transversemass}.\footnote{The spill-out of $t \bar t$ events beyond $m_T(\ell, \met) = m_W$ is caused by dileptonic events (with one lepton being lost) or by events with a $\tau$ (either leptonic or hadronic). Another possible source of these events can possibly be detector smearing, which is not captured here, since we do not run detector simulation. We expect that the latter effect is minor.} At higher stop masses, 3-body decays become less important, but tails of kinematic distributions in which the bino momenta are as large as possible play a key role in the enhanced efficiency. Similar results were presented for Tevatron searches in Fig.~9 of Ref.~\cite{Kats:2011it}, which focused on the gravitino case in which even more 3-body decays are present and the stop to top ratio can be even higher near the top mass. The lesson is that even in the stealth regime, missing $E_T$ variables, especially clean ones like $M_T$ or $M_{T2}$ that can be designed to have edges in the background, can still play a useful role in constraining stops. However, 30\% to 60\% gains in $S/B$ when starting with a signal an order of magnitude below the background leave much to be desired. A further difficulty is that, at these low masses, stop events will contaminate control regions of many studies, making it more difficult to infer the presence of a signal. We expect that in this regime missing energy variables are best used as part of a larger toolkit that exploits other differences in top and stop kinematics. In the rest of the paper, we will aim to build that toolkit.

Before continuing, let us make a simple estimate of the number of events that we can expect to be able to work with. With 20 fb$^{-1}$ and an NLO $t{\bar t}$ rate of about 200 pb, and estimating a typical selection efficiency of about 20\%, we expect to have (ignoring taus) about 40,000 dileptonic top events, 200,000 semileptonic top events, and 350,000 all-hadronic events. In this paper we will concentrate on signals with leptons, to avoid the potentially large QCD backgrounds in the all-hadronic channel. Note that if we knew the dileptonic top rate perfectly, a sample of 4,000 stops on top of 40,000 tops would have $S/\sqrt{B} \sim 20$. Of course, in reality systematic uncertainties overwhelm a simple cut-and-count method, but this number shows that our task is not hopeless, provided we can find shape variables with convincing differences.

%%%%%%%%%%%%%%%%%%%%%%%%%%%%%%%%%%%%%%%%%%%%%%%%%%%%%%%%%
%\section{Spinning Top: Angular Correlations}
\section{Spin Correlations}
\label{sec:spin}
%%%%%%%%%%%%%%%%%%%%%%%%%%%%%%%%%%%%%%%%%%%%%%%%%%%%%%%%%

\subsection{Spin correlation}
In the SM tops, being fermions, are pair produced with correlated polarizations that manifest in certain measurable angular variables~\cite{Jezabek:1988ja,Stelzer:1995gc,Mahlon:1997uc,Mahlon:1995zn,Brandenburg:1996df,Parke:1996pr,Mahlon:2010gw}.  For instance, at the LHC, where top production is dominated by $gg\rightarrow t\bar t$, when one top is left-handed the other tends to be as well, and similarly for the right-handed case (see \Ref{Parke:2010ud} for an overview, and~\cite{Baumgart:2011wk} for a study in the context of new physics).  However, tops produced via stop decays have no such correlation because the scalar stops can not preserve any spin information. Therefore we can take advantage of techniques that were previously developed to measure spin correlation in $t \bar t$ pairs. In our case the goal is much more challenging. We do not compare the hypothesis that the top and antitop have Standard Model spin correlation with the hypothesis that they are completely uncorrelated, but rather with the hypothesis that a spin-correlated $t \bar t $ sample has ${\cal O}(10\%)$ contamination from scalar events, which approximately look like spin-uncorrelated tops.\footnote{One other effect that could play a role in angular distributions turns out to be unimportant for us: the stop can be mostly right-handed or mostly left-handed (as some theoretical models predict; see e.g.~\cite{Csaki:2012fh}), and so the tops coming from the stop decays can be polarized. While it can be an appreciable effect if the mass splitting between top and stop is large~\cite{Perelstein:2008zt,Shelton:2008nq}, it is a small effect in the stealthy regime, as we have checked explicitly. Hence, we will not discuss it further.}

\FIGURE[ht]{
\includegraphics[width = 0.6\textwidth]{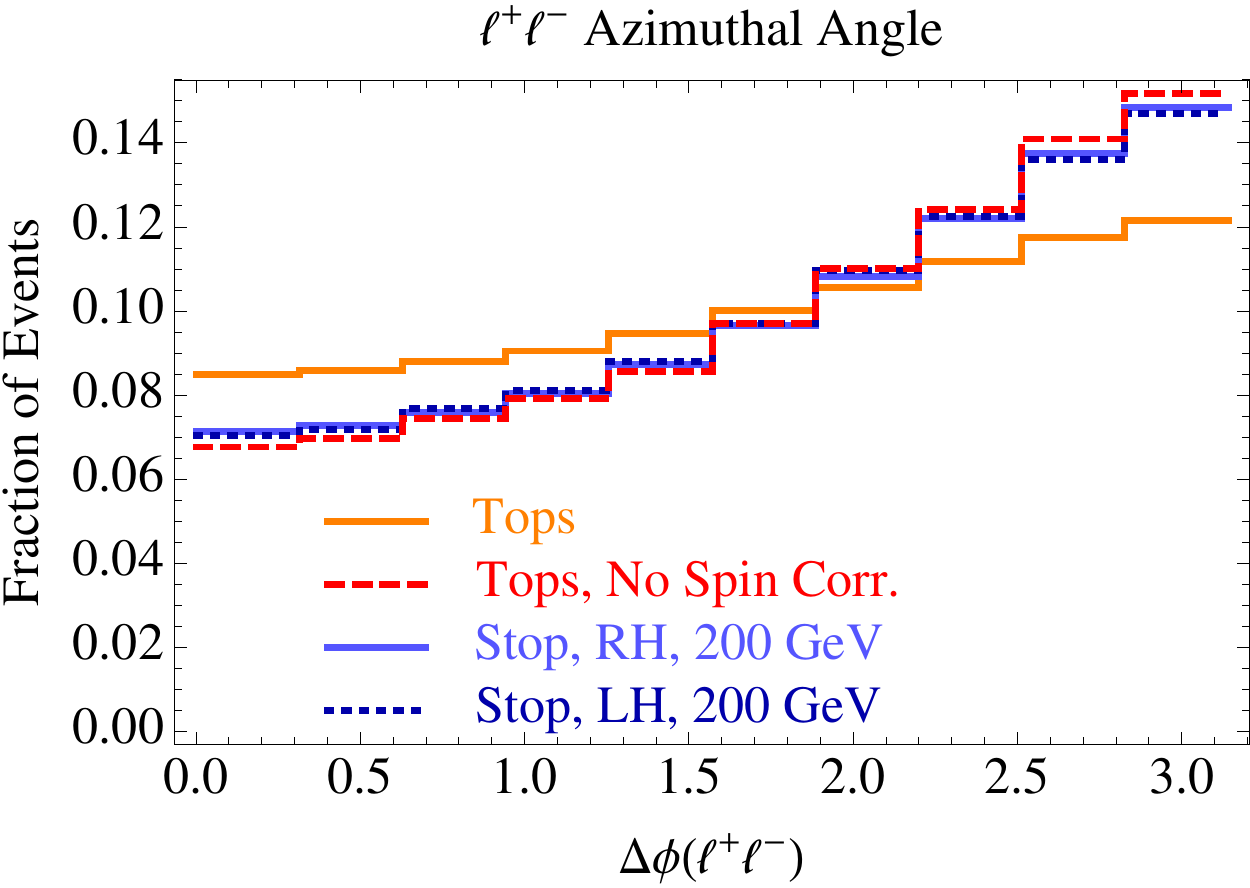}
\caption{$\Delta \phi(\ell^+,\ell^-)$ for $t{\bar t}$ production, ${\tilde t}{\tilde t}^\ast$ production, and $t{\bar t}$ production with spin correlation turned off (i.e., the differential rates for production and decay are factorized and we randomize the top helicities in between). Notice that, from the point of view of this variable, stops are essentially the same as spin-uncorrelated tops. Also, polarization effects are small, as left- and right-handed stops have the same distribution.}
\label{fig:stopspinuncorr}
}

%zhenyu
When the LSPs are soft, stop events are similar to top pair events without correlation. This is illustrated in Figure~\ref{fig:stopspinuncorr}, which shows one distribution, $\Delta \phi(\ell^+,\ell^-)$, which is sensitive to spin correlations, and for which stops look like tops with spin correlation turned off. We have calculated the observable for tops with MC@NLO~\cite{Frixione:2002ik,Frixione:2003ei} at parton level, and checked that corrections from varying the top mass and the renormalization and factorization scales are small relative to the shift that would arise from adding a sample of stops to the tops. This observable has been studied by ATLAS to probe the existence of spin correlations in $t{\bar t}$ production~\cite{ATLAS-CONF-2011-117}, with the most recent update achieving 5$\sigma$ significance for the existence of nonzero correlation effects~\cite{ATLAS:2012ao}.  

In order to confirm the SM top pair spin correlation  Ref.~\cite{Melnikov:2011ai} proposed a method using full matrix elements with and without spin correlation. This method has been implemented experimentally in Tevatron searches~\cite{Abazov:2011ka, Abazov:2011gi}, which observed  evidence for spin correlation in both the dileptonic and semileptonic channels. Since many more top events are produced at the LHC than at the Tevatron, we are expecting a more precise measurement at the LHC of the $t\bar t$ spin correlation. Any deviation from the SM prediction will be a sign of new physics. In the presence of light stops, we will observe a mixture of correlated and uncorrelated top pairs. In the following, we discuss the use of the matrix element method in stop searches. We concentrate on the dileptonic channel in the following discussion. We have also checked the semileptonic channel, which gives a less significant effect due to combinatorial uncertainties. 

For a dedicated stop search, it would be optimal to use directly the stop matrix element, which involves both the matrix element for the $2\rightarrow 2$ process and the one for subsequent particle decays. However, as we will discuss further in Section~\ref{sec:rapidity}, the differential cross section for the $2\rightarrow 2$ process is subject to large theoretical uncertainties. On the other hand, it was observed in Ref.~\cite{Melnikov:2011ai} that the spin correlation information for $t\bar t$ production is stable against NLO corrections at the LHC. Therefore, it is illuminating to examine the spin correlation alone, without invoking explicitly the stop pair matrix element. For this purpose, we use the matrix elements for top pairs with and without correlations, as they were given in~\cite{Mahlon:2010gw}. (The spin uncorrelated matrix element was derived from the assumption that top decays are completely spherically symmetric in the rest frame of the top). This is also a generic approach that allows us to find signs of new physics in $t\bar t$ correlations without specifying the underlying theory. 

\subsection{Implementation in stop searches}
In the dileptonic channel, we reconstruct the events and calculate the likelihood distributions as follows. Top and stop events are generated at parton level (including spin correlations) with MadGraph~\cite{Alwall:2011uj} and showered with Pythia~\cite{Sjostrand:2006za,Sjostrand:2007gs}. Jets are clustered with FastJet using the anti-$k_T$ algorithm with radius 0.4~\cite{Cacciari:2005hq,Cacciari:2011ma}, and we apply simple isolation requirements to define leptons, which are identical to the isolation criteria of~\cite{ATLAS:2012aa}.
\begin{enumerate}
\item Events are required to pass the following kinematic cuts, which closely follow (but are not identical to) the event selection used by ATLAS in Ref.~\cite{ATLAS:2012aa}.
\begin{itemize}
\item At least two jets with $p_T>25\gev$ and $|\eta|<2.5$, one or two of which must be b-tagged. 
\item Two opposite-sign leptons with $p_T>20\gev$ and $|\eta|<2.5$.
\item In the $e^+e^-$ and $\mu^+\mu^-$ channels, the invariant mass of the two leptons must satisfy $m_{\ell\ell} > 15\gev$ and $|m_{\ell\ell} - M_Z|< 10\gev$. In addition we demand $\met > 40\gev$.
\item In the $e\mu$ channel, no cut is applied on $m_{\ell\ell}$ or $\met$. Instead, a cut on $H_T$ (defined as the scalar sum of all selected charged leptons and jets) is applied: $H_T>130\gev$. 
\end{itemize}
\item We apply a flat b-tagging efficiency of 65\% (this is conservative relative to the 80\% quoted in Ref.~\cite{ATLAS:2012aa}). For events with only one b-jet, we consider all non-b jets with $p_T>25\gev$ as a candidate for the other b-jet. Then for each combination of two b-jets and two leptons, we can use the $W$ and top mass shell constraints to solve for the momenta of the two neutrinos from $W$ decays, assuming they are the only missing particles in the event. There could be 0, 2 or 4 real solutions. We discard events without real solutions. The final efficiency is 16.8\% for tops and 15.6\% for stops. All solutions from all combinations are taken into account in the following calculation. Following Ref.~\cite{Melnikov:2011ai}, we define the probability distribution for a given event as 
\begin{equation}
P_H=\mathcal{N}_{H}^{-1}\sum_{ij}\sum_{a}J_a f_i^{(a)}f_j^{(a)}\left|\mathcal{M}_H^{ij}\left(p_{\text{obs}}, p_\nu^{(a)}, p_{\bar\nu}^{(a)}\right)\right|^2,
\end{equation}
where $H= \{\text{corr,uncorr}\}$ denotes the hypothesis of correlated or uncorrelated tops, $\mathcal{N}_{H}$ is a normalization factor, $J_a$ is the Jacobian which appears when integrating over the neutrino momenta, $f_i$ and $f_j$ are the parton distribution functions for the two incoming partons, and $\mathcal{M}_H^{ij}$ is the leading order matrix element. The sum over $a$ is the sum of all combinations and solutions. The sum over ${i,j}$ is over all possible initial partons, $gg$, $u\bar u$ and $d\bar d$. For each event, we calculate both $P_{\rm corr}$ and $P_{\rm uncorr}$ and define the variable $\mathcal{R}$ as
\begin{equation}
\mathcal{R}= \frac{P_{\rm corr}}{P_{\rm corr}+P_{\rm uncorr}}.
\end{equation}   
The variable $\mathcal{R}$ is the likelihood for an event to be a correlated top pair. The ${\cal R}$ distributions for top event and stop events are shown in Fig.~\ref{fig:R_dis}. For comparison, we also show the ${\cal R}$ distribution at the parton level.
\FIGURE[ht]{
%\vspace*{3mm} 
\includegraphics[width=\textwidth]{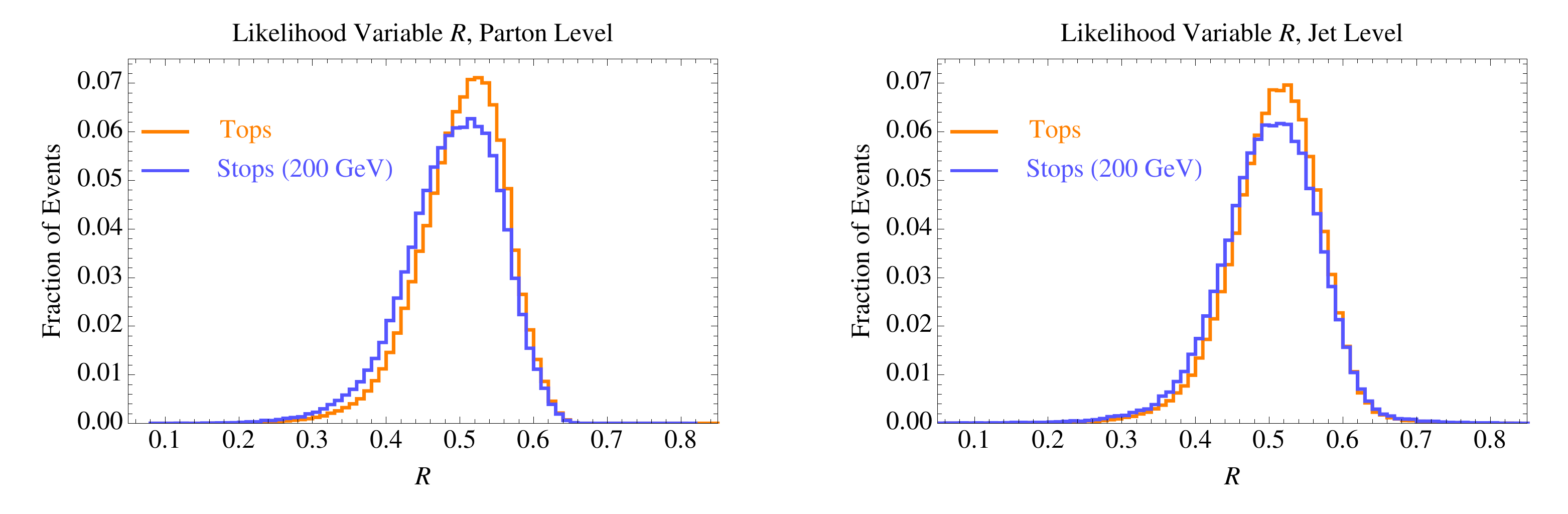}
\caption{The likelihood variable ${\cal R}$. Left: parton level; right: jet level.}
\label{fig:R_dis}
}

\item Given the ${\cal R}$ distributions, we can also follow Ref.~\cite{Melnikov:2011ai} to calculate the log likelihood ratio $L$ to discriminate between two hypotheses with a set of $N$ events. Here, $L$ is defined as
\begin{equation}
L=2\text{ln}(\mathcal{L}_t/\mathcal{L}_{\tilde t}),
\end{equation}
where 
\begin{equation}
\mathcal{L_K}\equiv\prod_i^N\rho_K(\mathcal{R}_i)
\end{equation} 
and $\rho_K(\mathcal{R}_i)$ is the probability density read from Fig.~\ref{fig:R_dis}.
 we choose the number of top pair events corresponding to 20 $\text{fb}^{-1}$ at 8TeV and compare it with the same number of events, but with top and stop mixed in the ratio of 12:1, corresponding to the ratio in the cross sections. After kinematic cuts and reconstruction, we are left with 32.8k top events for the pure samples, and 30.4k top events and 2.4k stop events for the mixed sample. We generate 10k pseudo-experiments corresponding to the two cases and the resulting $L$ distributions are shown in Fig.~\ref{fig:L_dis}. For comparison, we also generate 10k pseudo-experiments at the parton level using the same number of events as the jet level (after reconstruction). From the jet level result in Fig.~\ref{fig:L_dis}, we estimate that on average a light stop of $200\gev$ can be excluded at 95\% confidence level.  
  
 \FIGURE[ht]{
%\vspace*{3mm}
\includegraphics[width=\textwidth]{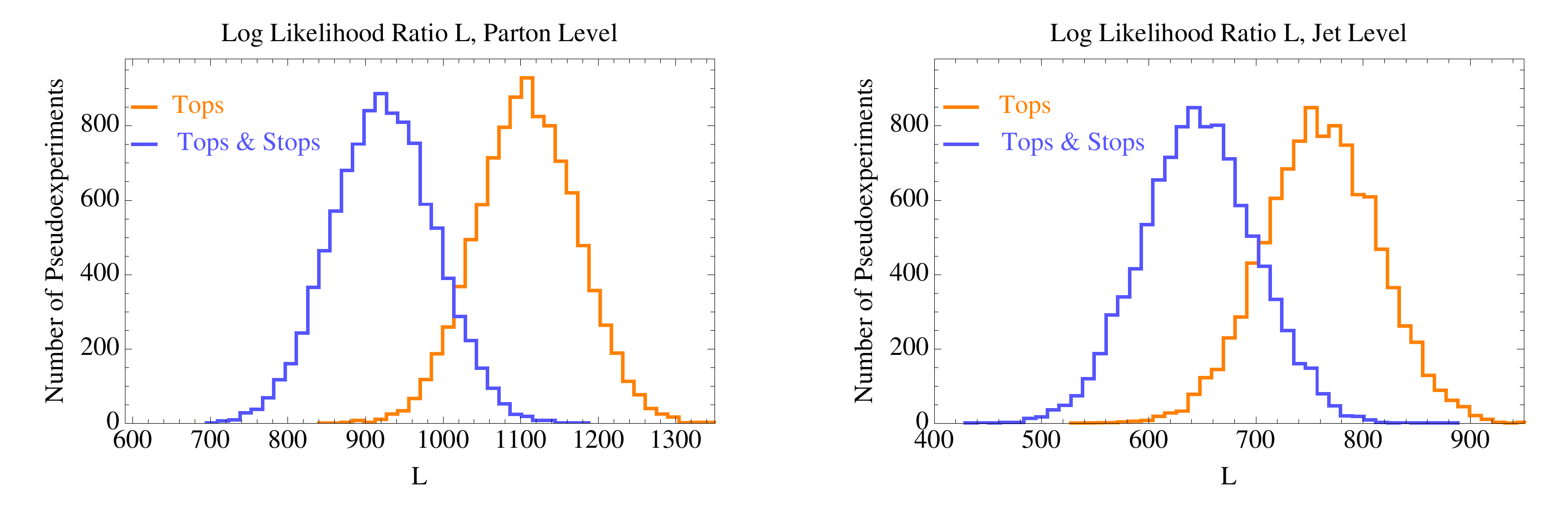}  
\caption{The log likelihood ratio $L$. Each point on the curve corresponds to a pseudoexperiment with 32.8k events. Note in particular that each top plus stop pseudoexperiment is normalized to the same number of events as each top pseudoexperiment. Left: parton level; right: jet level.}
\label{fig:L_dis}
}

\end{enumerate}

%-zhenyu

%%%%%%%%%%%%%%%%%%%%%%%%%%%%%%%%%%%%%%%%%%%%%%%%%%%%%%%%%
%\section{Stop, Look, and Listen: Rapidity Gaps}
\section{Rapidity Gaps}
\label{sec:rapidity}
%%%%%%%%%%%%%%%%%%%%%%%%%%%%%%%%%%%%%%%%%%%%%%%%%%%%%%%%%

\subsection{Top and stop production amplitudes}

\begin{table}[ht]
\centering
\large
\begin{tabular}{l|ll}
Initial state & $t{\bar t}$ & ${\tilde t} {\tilde t}^\ast$ \\
\hline
$gg$ & 68 pb & 11 pb\\
$q{\bar q}$ & 23 pb & 1.6 pb \\
\end{tabular}
\caption{LO cross sections for top and stop production processes in MadGraph, for a 180 GeV stop. Notice, in particular, the tininess of stop production from $q{\bar q}$ due to the $p$-wave threshold, as well as the significant suppression of stop relative to top production even from gluons.}
\label{tab:topstoprates}
\end{table}

As illustrated in Figure~\ref{fig:stopcrosssection}, the pair production rate of stops, even with a mass equal to the top mass, is well below the pair production rate of tops. Delving into the differences in stop and top production processes will shed some light on this, and also suggest a set of variables that can be useful in discriminating top from stop events. The rates for top and stop production at leading order, from either $q{\bar q}$ or $gg$ initial states, are shown in Table~\ref{tab:topstoprates}. The most striking fact is the smallness of $q{\bar q} \to {\tilde t} {\tilde t}^\ast$, which is explained by $p$-wave suppression: the stops in the final state need to carry angular momentum. Since they have no spin, this implies that they are produced in a $p$-wave with a rate $\propto \beta^3$ near threshold. However, no such simple argument explains the ratio between stop and top production initiated by gluons. Naively, we might expect that the top and stop are both color triplets, so that for energies far above threshold, their production rates will be identical up to factors counting degrees of freedom, so that the rate for stops is asymptotically half that of tops. Table~\ref{tab:topstoprates} makes it clear that this misses some aspect of the physics.

We can get some insight by considering the behavior of the differential cross section in the massless limit. For stops, there is a well-defined total rate~\cite{Beenakker:2010nq}:
\begin{equation}
\sigma(gg\to {\tilde t}_1  {\tilde t}^\ast_1) \to_{s\gg m} \frac{5\alpha_s^2\pi}{48 s},
\end{equation}
On the other hand, for massless quarks, one has~\cite{Combridge:1977dm}:
\begin{equation}
\frac{d\sigma}{d\Omega}(gg \to q\bar q) = \frac{\alpha_s^2}{24 s} \left(t^2 + u^2\right) \left(\frac{1}{t u} - \frac{9}{4s^2}\right),
\end{equation}
which doesn't have a well-defined integral over phase space due to the $t$- and $u$-channel poles. Because the top is massive, we expect that this integral will be regulated and that the result will be a $\log\frac{s}{m_t^2}$ enhancement in the rate for forward top production in the parton center-of-momentum frame. This is a very real physical difference between the production of top quarks and scalar top quarks, which we should try to exploit. It is easy to see by considering the even simpler case of scalar QED versus QED, where the simplest MHV amplitudes with the right little group properties are easily written down and give the correct answers:
\begin{eqnarray}
A^{\rm tree}(1^+,2^-,3_\phi, 4_\phi) & = & ie^2 \frac{\left[1~3\right]\left<2~3\right>}{\left<1~3\right>\left[2~3\right]} = e^2 \times~{\rm phase} \\
A^{\rm tree}(1^+,2^-,3^-_{\bar \psi},4^+_\psi) & = & ie^2 \frac{\left[1~4\right]\left<2~3\right>}{\left<1~3\right>\left[2~3\right]} = e^2\sqrt{\frac{u}{t}}\times~{\rm phase}.
\end{eqnarray}
The latter case corresponds to the familiar splitting function ameliorating the pole in a $t$-channel diagram to the square root of a pole in the amplitude. In the scalar case, this pole is completely absent. The two results are, in fact, related by a SUSY Ward identity.

\subsection{Parton-level distributions and systematics}

\FIGURE[ht]{
\includegraphics[scale=0.6]{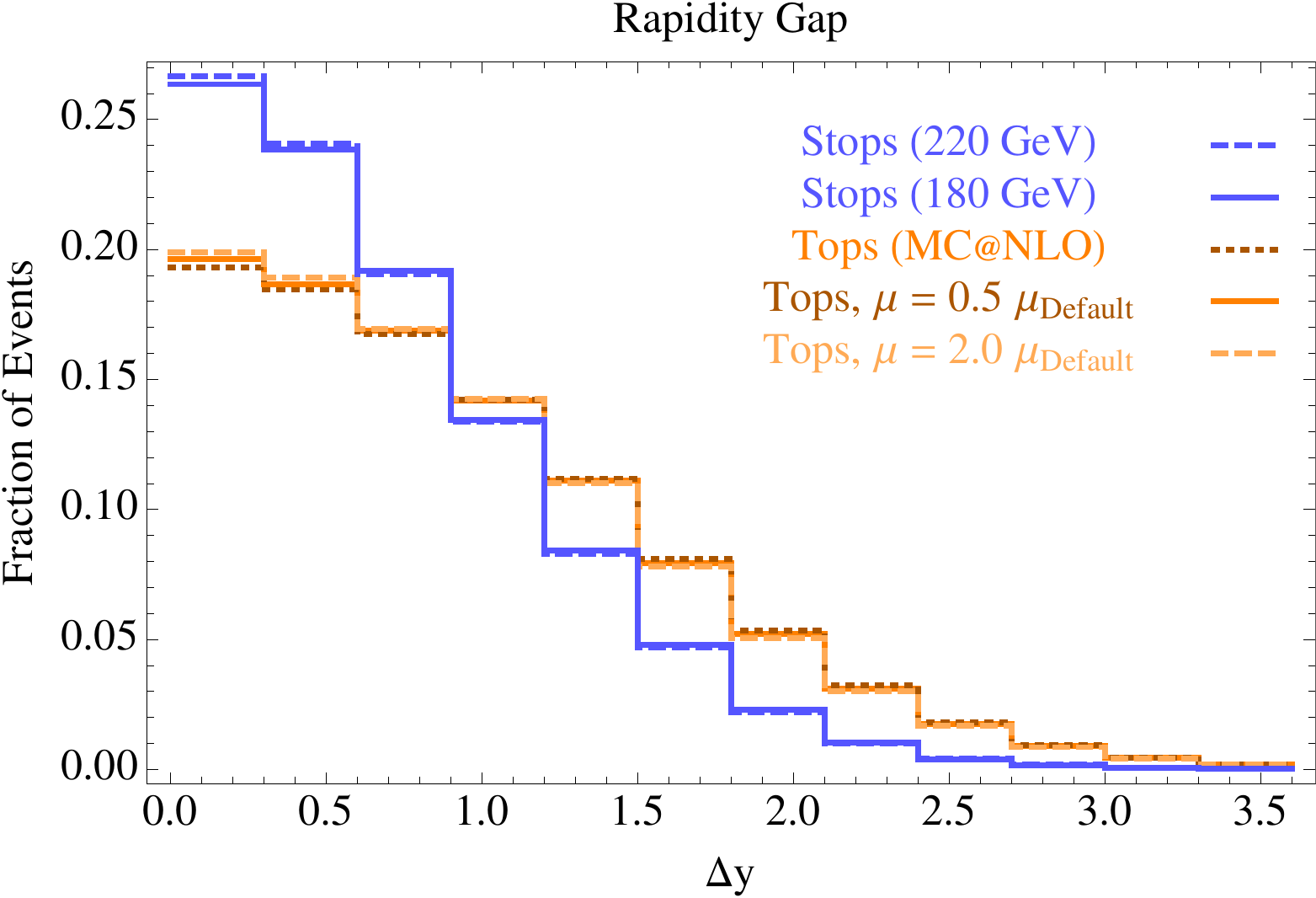}
\caption{$\Delta y$ distribution between top and antitop for $t{\bar t}$ production (orange) and for a 180 GeV scalar top (blue), both normalized to 1. Stops are more likely to be produced at small rapidity difference because the $t$- and $u$-channel poles in fermionic quark production from $gg$ are absent in scalar quark production. Three curves are shown for tops, corresponding to different choices of renormalization and factorization scales in MC@NLO.}
\label{fig:rapiditydiff}
}

Our intuition from the limits of massless particles is useless if it does not carry over to a fact about physical, massive tops and stops, but it does. The distribution of rapidity gaps between produced tops and stops is shown in Figure~\ref{fig:rapiditydiff}. The distributions are clearly different. However, we should keep in mind the fundamental fact about stop rates: in a sample of candidate top events, we are usually looking for at most 10\% of the sample composed of stops, so even a 50\% difference in the shape of pure top and stop distributions will become a 5\% effect in the combined sample. We must investigate the robustness of our Monte Carlo predictions of the shape of the top rapidity gap distribution, to understand whether such an effect can ever be measured. We will begin by assessing this at parton level, ignoring subtleties associate with jets and detectors. Making the case that the difference is observable at this level is a necessary but not sufficient precondition for claiming that it can be measured at the LHC.

We have generated large samples of $t{\bar t}$ events using MC@NLO~\cite{Frixione:2002ik,Frixione:2003ei} to assess the systematic uncertainties. We have varied both the factorization and renormalization scales, and the top quark mass, to analyze the effects on the shape of the $\Delta y$ distribution. The fractional change in shape, bin-by-bin, is plotted in Figure~\ref{fig:scalevsstop}. Here we have computed the binned $\Delta y$ shape for different scale choices, normalized the shapes to unit area, and given the size of the shift in each bin relative to MC@NLO with default choices. We do not show variations of the top mass in the plot because we found the shape to be insensitive to it. We also show in Figure~\ref{fig:scalevsstop} the change in the default shape from tops when 12\% of the sample is composed of 180 GeV stops instead of tops. Although the shape change can be significant, we see that much of the effect can be mimicked by increasing the renormalization scale. This compares only the {\em shape}, not the normalization. Doubling the renormalization scale also lowers the cross section by about 13\%, whereas the presence of 180 GeV stops increases it by about 12\%. However, the lowered cross section can be compensated to some extent if the top mass is lower, and the total top rate has at least a 10\% theoretical uncertainty, as we reviewed in Section~\ref{sec:current}. The bottom line is that the rapidity distribution carries definite physical information, but to use it we must be cautious about systematic uncertainties in our understanding of top quark production. Continuing progress in understanding of tops at NNLO~\cite{Baernreuther:2012ws} could play a role in reducing these uncertainties. 

Parton distribution functions could also be a source of uncertainty. We have not systematically explored this, but as one illustrative example, switching MC@NLO's choice of PDF set from the default CTEQ 6M to an Alekhin NLO FFN set produced a curve similar to the $\mu = 0.5~\mu_{\rm Default}$ curve in Figure~\ref{fig:scalevsstop}. Understanding PDF uncertainties would clearly be one component of getting Standard Model $t{\bar t}$ predictions under enough control to make statements about the presence of new physics with any confidence.

\FIGURE[ht]{
\includegraphics[scale=0.6]{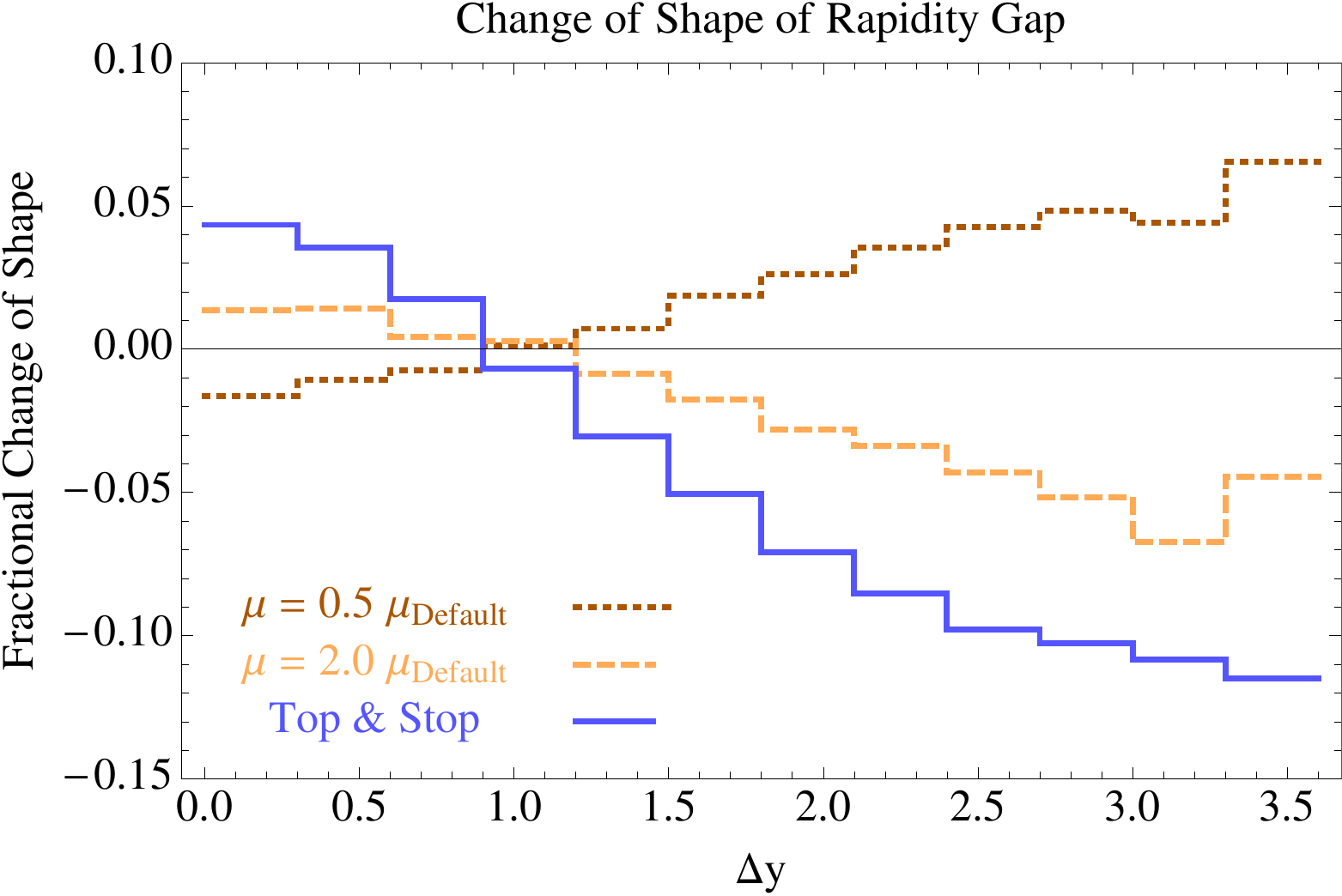}
\caption{Relative fractional change in the fraction of events in each bin of $\Delta y$. The comparison is to MC@NLO run with default renormalization and factorization scales. The dotted dark orange curve is the result of halving these scales. The dashed light orange curve doubles the scales. The blue curve is the result of adding an admixture of 180 GeV stops (12\% of the events, based on the relative NLO cross sections). The concern is that the effect of stops may be mimicked by a larger renormalization scale.}
\label{fig:scalevsstop}
}

So far we have been discussing only inclusive parton-level quantities. Cuts on kinematic variables can help to draw out the differences between stops and tops. In Figure~\ref{fig:rapiditydiff}, both stops and tops peak at $\Delta y = 0$. This is because both are dominantly produced near threshold, where one has $\Delta y = 0$ by definition. In fact, one can easily see that as a function of the center-of-mass energy of the parton collision, $\sqrt{\hat s}$, there is a bound on $\Delta y$ in $t{\bar t}$ events:
\begin{equation}
\Delta y \leq \log \frac{{\hat s} + \sqrt{{\hat s} - 4 m_t^2}}{{\hat s} - \sqrt{{\hat s} - 4 m_t^2}}
\label{eq:deltaymax}
\end{equation}
The same consideration applies to the $t{\bar t}$ subsystem of a stop event. Figure~\ref{fig:mtt_vs_deltay} shows where events lie in the plane of $(M_{t\bar t},\Delta y)$; tops are seen to prefer larger $\Delta y$ at larger $M_{t\bar t}$. This is not surprising since the $t$-/$u$-channel singularity is regularized by $m_t$, and as $\hat s$ grows, tops approaches a massless quark limit. Thus, a cut on $\hat s$ will allow us to isolate the regime where there {\em can}, in principle, be large differences between top and stop events. However, there is a caveat: a hard cut on $\hat s$ will make the difference in distributions clear, but runs the risk of diluting our samples enough that statistical uncertainties overwhelm the systematics and prevent us from drawing a clean conclusion. 

\FIGURE[ht]{
\includegraphics[width=\textwidth]{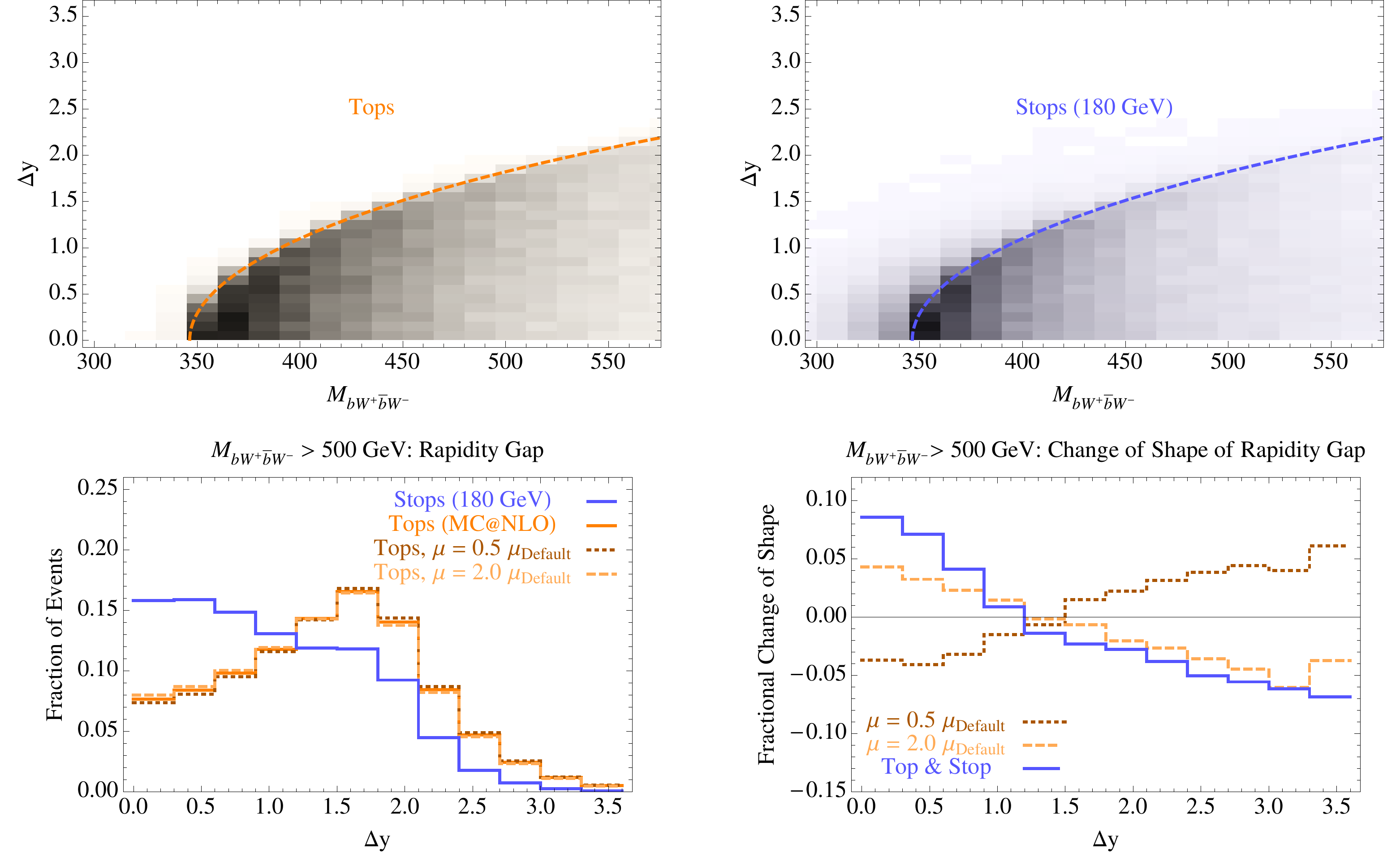}
\caption{{\bf Top}: correlation between the $t{\bar t}$ invariant mass (which for tops characterizes the center-of-mass energy $\sqrt{\hat s}$ of the parton collision) and the value of $\Delta y$. Notice that at larger $M_{t{\bar t}}$, the tops are more strongly peaked at larger $\Delta y$ than the stops. The dashed curves correspond to Eq.~\ref{eq:deltaymax}; some stop events spill over the line, because the decays are off-shell (in which case we have replaced $p_{\rm top}$ with $p_b + p_W$ in the definitions of $M_{t{\bar t}}$ and $\Delta y$).  {\bf Bottom}: the analogues of Figures~\ref{fig:rapiditydiff} and~\ref{fig:scalevsstop} after imposing a cut $M_{t\bar t} > 500$ GeV. The mass cut leads to a more dramatic shape difference, but the systematic effect of varying the renormalization and factorization scales can still mimic the signal.}
\label{fig:mtt_vs_deltay}
}

As the bottom right-hand plot of Figure~\ref{fig:mtt_vs_deltay} shows, the mass cut also does not eliminate the systematic uncertainty: varying the factorization and renormalization scales upward by a factor of 2 continues to mimic much of the effect of adding stops to the top sample. It may be that a variable can be constructed that will capture similar physics while reducing systematic uncertainties. We have studied the Collins-Soper angle, which corrects the angle of the top quark to the beam direction to minimize the effects of ISR~\cite{Collins:1977iv}, as one possibility, but found that it was not an improvement.

\subsection{Jet-level distributions and practical use}
\label{subsec:jetlevel}

In spite of the systematic uncertainties discussed in the previous subsection, 
it can be useful to exploit the 
rapidity gap as one more variable that can potentially discriminate between 
a pure top sample and a stop-enriched one.
The most straightforward way would be reconstructing the entire event and using 
the rapidity gap between the tops. Although possible, we find that this approach 
is not ideal. It is clear that 
some stop events, where the neutralinos contribute an appreciable portion of the 
$\met $, will not 
pass reconstruction criteria. In this way, we lose the most 
pronounced non-top-like events. Nonetheless, applying the same reconstruction as in Section~\ref{sec:spin}, we find $2\,\sigma$ significance from $\Delta y(t,\bar t)$ alone. (The combination with spin correlations is discussed in Appendix~\ref{sec:multivariate}.)

To avoid this 
undesired loss of valuable events, we find another variable which is closely tied 
to the 
rapidity gap between the tops. It turns 
out that the rapidity gap between the leptons in the dileptonic events 
closely follows the rapidity gap 
between the tops. The rapidity gap between the leptons also grows in 
a $t\bar t$ sample as $m_{t \bar t}$
increases.

We also do not use the top invariant mass in our study, because it would 
force us again to reconstruct the event. 
One might also consider
reconstructing the invariant mass of \emph{all} the objects in the event but this  would 
introduce an undesired
sensitivity to the ISR/FSR, deteriorating the correlation with $m_{t\bar t}$.
Instead
we choose a different quantity which closely mimics the behavior of the 
invariant mass and can be easily
calculated without full event reconstruction. This quantity is an invariant 
mass of the leading visible objects 
in a dileptonic $t\bar t$ event. If both b-jets are tagged, we simply 
construct an invariant mass of 
two leptons and two b-jets. If only one b-jet is tagged, we construct an invariant 
mass of the leptons, 
b-jet and the non-tagged jet with the highest $p_T$ in the event.  

\FIGURE[ht]{
\includegraphics[width=1.0\textwidth]{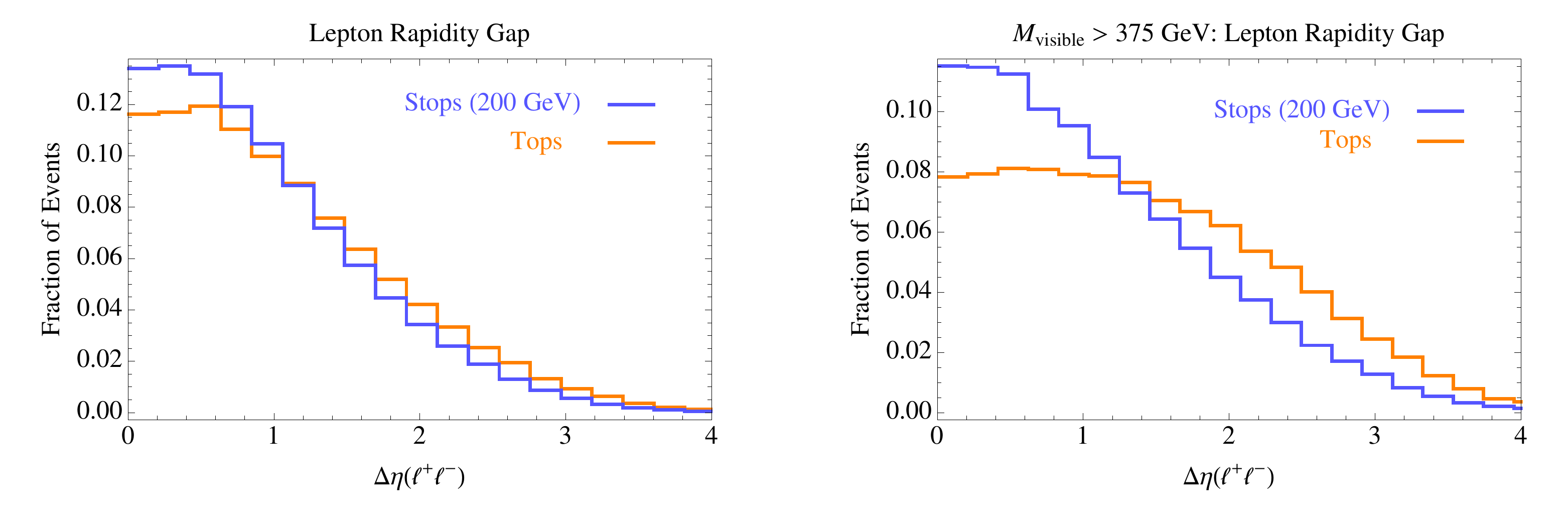}
\caption{Rapidity gap distribution for tops and stops, normalized to the rate of 
events passing the cuts. 
The distribution on the LH side includes all the events which pass simple 
$t\bar t$ acceptance cuts. 
The distribution on the right shows only those events which have a visible 
mass higher than 375 GeV, as explained
in the text.}
\label{fig:deltaynormal}
}

\FIGURE[ht]{
\includegraphics[width=0.6\textwidth]{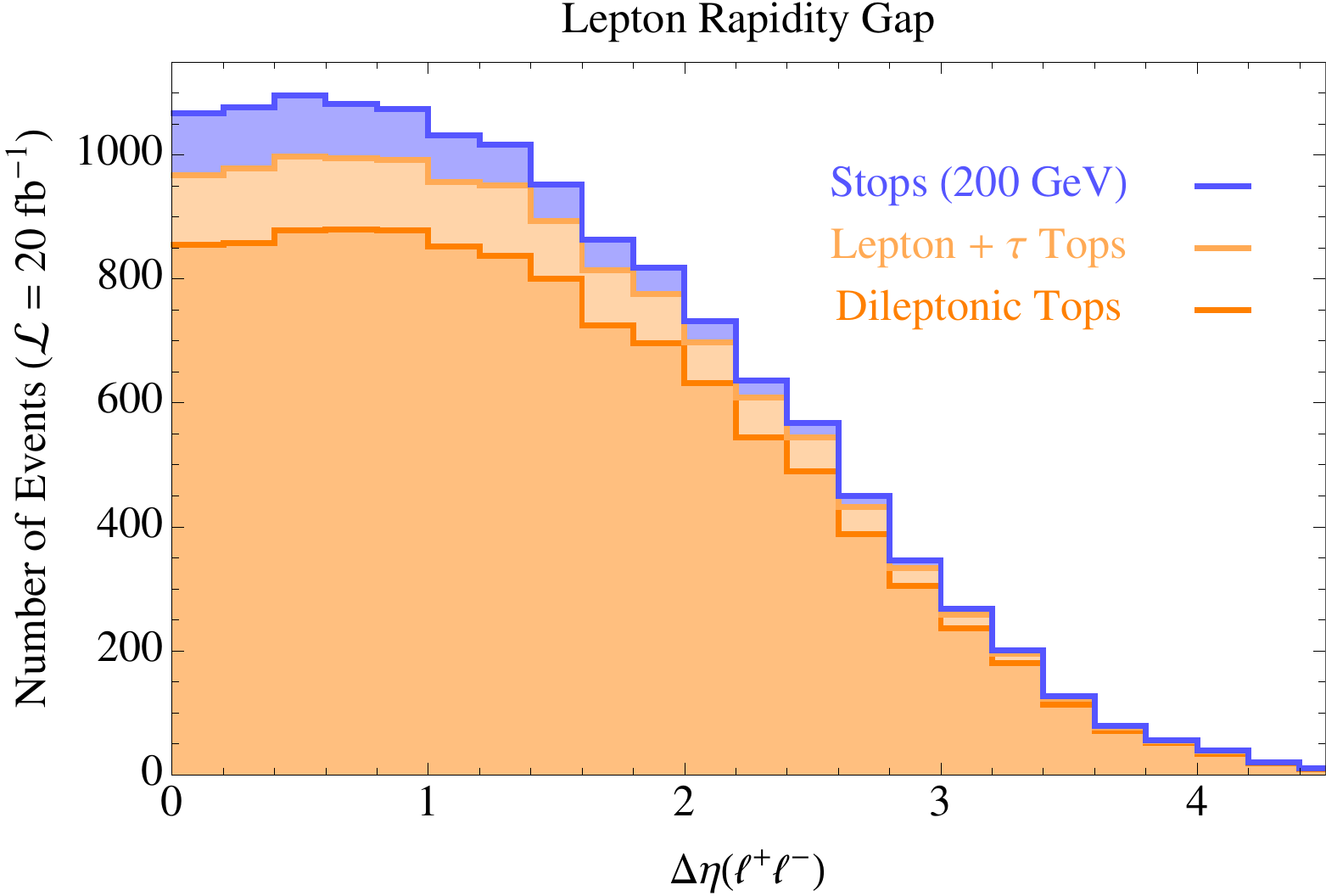}
\caption{Rapidity gap between the leptons after the cut on the ``visible mass'' as explained in the text.}
\label{fig:stucked_eta}
}
            
We illustrate our approach in Fig.~\ref{fig:deltaynormal}. To build these plots we select the following events (these selection criteria are very similar to~\cite{CMS-PAS-TOP-11-005,ATLAS:2012aa}):
\bi
\item Events with precisely two isolated leptons. The $p_T$ of each lepton must exceed 20 GeV and 
$|\eta_l| < 2.5$.\footnote{Even softer isolated muons were allowed in the search~\cite{CMS-PAS-TOP-11-005}, if they were the softest lepton in the event, so in this sense our results can be further improved. }
\item Each event should have two or more jets, $p_T (j) > 25$ GeV, $|\eta(j)| < 3.0$.
\item At least one jet should be b-tagged; we assume a flat tagging rate of 65\%.
\item $\met > 35$ GeV
\item Discard the events with OSSF leptons and invariant mass $76\ {\rm GeV} < m_{ll} <
106 \ {\rm GeV}$ ($Z$-window)
\item In the second sample we demand the invariant mass of the leading visible 
objects as explained before 
to be higher than 375 GeV.     
\ei 

We also plot in Fig.~\ref{fig:stucked_eta} an event distribution in a mixed top-stop sample and compare 
it to pure top. Notice that our cuts slightly favor stops. The acceptance 
for stops is 26.5\%, while the acceptance for tops is 25.5\%. This happens because the 
stop sample tends to have slightly higher $\met $ than tops, even though this tendency 
is not sufficient to discriminate between them alone. This is the feature we illustrated early in the paper in Fig.~\ref{fig:metnorman}. 
Note that the excess of events in in the bins with 
$|\Delta \eta | < 1$ exceeds 10\%, and maybe in combination with other distributions 
can be a good variable to discover or exclude the light stop. 

Finally we briefly discuss (lack of) contamination of these samples with other 
backgrounds. First, one might worry about $(Z \to \tau_\ell \tau_\ell)+ jj$, but 
it was shown in~\cite{CMS-PAS-TOP-11-005} that this background becomes negligible when two or
more jets are required. The second important background is dileptonic DY production 
with jets. In this background the missing $E_T$ usually comes from mismeasurement. 
It was also shown in~\cite{CMS-PAS-TOP-11-005} that a modest cut on $\met > 35$ GeV and the requirement 
of two jets render this background subdominant. Although this search was performed 
for $\sqrt{s} = 7$ TeV, there is no reason to believe that the results of a $\sqrt{s} = 
8$ TeV search will be very different. Hence, have we neglected these 
backgrounds throughout the paper.   

%%%%%%%%%%%%%%%%%%%%%%%%%%%%%%%%%%%%%%%%%%%%%%%%%%%%%%%%%
%\section{Don't Stop Believin': Discussion}
\section{Discussion}
\label{sec:discussion}
%%%%%%%%%%%%%%%%%%%%%%%%%%%%%%%%%%%%%%%%%%%%%%%%%%%%%%%%%

Naturalness could still manifest itself through a variety of possible signatures. A heavy stop, for instance, could decay to a lighter stop to produce  $t{\bar t}Z+\met$~\cite{Perelstein:2007nx}. The sbottom provides its own novel and possibly easier-to-find signatures~\cite{Li:2010zv,Datta:2011ef, Lee:2012sy, Alvarez:2012wf}.  Higgsinos must be light for tree-level naturalness, and if they are the LSP or NLSP they can lead to more easily-observed stop decays~\cite{Papucci:2011wy,Asano:2010ut,Aad:2012cz}. Furthermore, deviations in Higgs boson production and decay rates from the Standard Model could also provide hints that naturalness is at work. Simultaneous pursuit of all of these signatures is necessary for a complete picture. 

Here we have focused on the ``stealth stop" regime in which missing momentum is relatively small, and argued that rapidity gap and spin correlation observables can help to build the case for a new physics signal. For less stealthy stops, we expect variants on missing $E_T$ (especially transverse mass variables and their generalizations, which are bounded in $t{\bar t}$ background) to be relatively successful at digging stop signals out of top backgrounds. Even in that case, however, we would emphasize that the rapidity gap and spin correlation observables are diagnostic of the production of scalar particles. They may be less essential for discovering a signal, but could play a key role in establishing the nature of the signal. In particular, the distinct rapidity distributions of stops as opposed to fermionic top partners gives a new handle on spin determination.  These measurements could be made with the next year's worth of data, and so we eagerly await the LHC's verdict on light stops.

\acknowledgments
We thank Lisa Randall for collaboration in the early stages of this project. We thank Rick Cavanaugh, Kirill Melnikov, Jessie Shelton, Matt Schwartz, and Brock Tweedie for useful discussions. ZH is supported in part by NSF grant PHY-0804450. AK is supported by NSF grant PHY-0855591. MR is supported by the Fundamental Laws Initiative of the Harvard Center for the Fundamental Laws of Nature. DK and MR thank the New Physics @ Korea Institute and the Shilla Seoul for hosting a productive workshop where a portion of this work was completed. The computations in this paper were run on the Odyssey cluster supported by the FAS Sciences Division Research Computing Group at Harvard University.

%%%%%%%%%%%%%%%%%%%%%%%%%%%%%%%%%%%%%%%%%%%%%%%%%%%%%%%%%
\appendix

%%%%%%%%%%%%%%%%%%%%%%%%%%%%%%%%%%%%%%%%%%%%%%%%%%%%%%%%%
%\section{I Want the World to Stop: Multivariate Analyses}
\section{Multivariate Analyses}
\label{sec:multivariate}
%%%%%%%%%%%%%%%%%%%%%%%%%%%%%%%%%%%%%%%%%%%%%%%%%%%%%%%%%

We have seen that there are various differences in distributions in top and stop events that can be understood on physical grounds, mostly related to the angular momentum properties of the final states. Although one could try to construct a stop search using any one of them, or using various cuts designed to enhance signal to background, we expect that the most flexible and powerful approach is a multivariate analysis that uses all of the information. First, we would like to check that the variables we have used so far are not highly correlated. 

\FIGURE[ht]{
\includegraphics[width=0.6\textwidth]{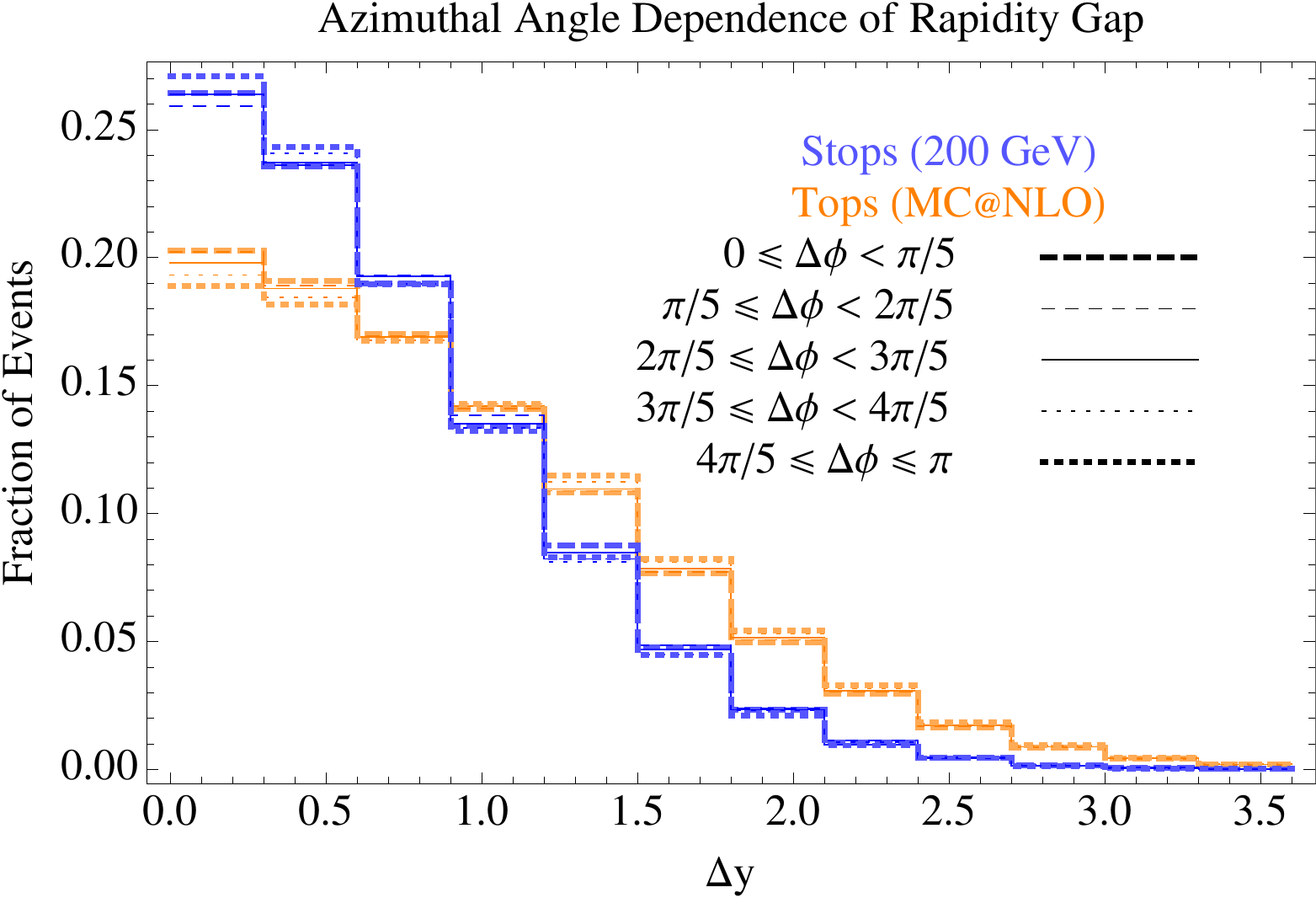}
\caption{Rapidity gap between top and antitop, as in Figure~\ref{fig:rapiditydiff}, with the samples binned based on $\Delta \phi(\ell^+,\ell^-)$. We see that the distributions are not strongly correlated, in the sense that restricting to a subset of $\Delta \phi$ values makes little difference in the shape of the $\Delta y(t,{\bar t})$ curve.}
\label{fig:corrcheck}
}

One way to check for correlations is to bin one variable and plot the other; we take this approach in Figure~\ref{fig:corrcheck}, which shows that selecting particular ranges of $\Delta \phi(\ell^+,\ell^-)$ makes relatively minor changes in the shape of the $\Delta y(t, {\bar t})$ distribution. Another check is to compare the two-dimensional distribution of simulated points in the $(\Delta y, \Delta \phi)$ plane with the product of two one-dimensional distributions. Comparing binned samples of $t{\bar t}$ events (10 bins in $\Delta \phi$ by 12 bins in $\Delta y$), we find that in any given bin, the ratio of the two-dimensional probability to the product of the two one-dimensional probabilities is always in the range 0.92 to 1.08. Hence, we expect that the discussions in Sec.~\ref{sec:spin} and Sec.~\ref{sec:rapidity} are approximately orthogonal, and we can achieve better significance by combining the two approaches. (Of course, one would want to be sure systematic uncertainties on $\Delta y$ are under control before taking this approach too seriously.)

Along these lines, we consider a simple multivariate method by taking the product of probabilities defined from one dimensional distributions \cite{Aaltonen:2010jr, Bai:2011uk}. For a given variable $x$, we obtain the signal and background distributions in two histograms with the same binning and normalized to the same area. Then we define the probability of an event in bin $i$ being a signal event as
\begin{equation}
p_s^x=\frac{n_s^i}{n_s^i+n_b^i}.
\end{equation}
Correspondingly, the event has a probability $(1-p_s^x)$ being a background event. For multiple variables, the total probability is defined as
\begin{equation}
\mathcal{P}_s=\frac{\prod_xp_s^x}{\prod_xp_s^x+\prod_x(1-p_s^x)}.
\end{equation} 

Then we can treat $\mathcal{P}_s$ as a single variable and repeat the likelihood test as in Sec.~\ref{sec:spin} for the two hypotheses and obtain the $L$ distributions. In Fig.~\ref{fig:L_combined}, we show the $L$ distributions by combining the spin correlation variable $\mathcal{R}$ and the reconstructed rapidity gap $\Delta y(t\bar t)$, for $20~\text{fb}^{-1}$ data. The reconstruction procedure is the same as in Sec.~\ref{sec:spin}. Note that in this approach, the correlations among different variables are not taken into account. Nonetheless, as for the $(\Delta y, \Delta \phi)$ distributions, the correlations between the two variables are small. Therefore we obtain significantly better discriminating power than using each individual variable: the two distributions are separated by $\sim 3\sigma$ in Fig.~\ref{fig:L_combined}, while the two variables alone each give us $\sim 2\sigma$ significance.  
 
\FIGURE[ht]{
\includegraphics[width=0.7\textwidth]{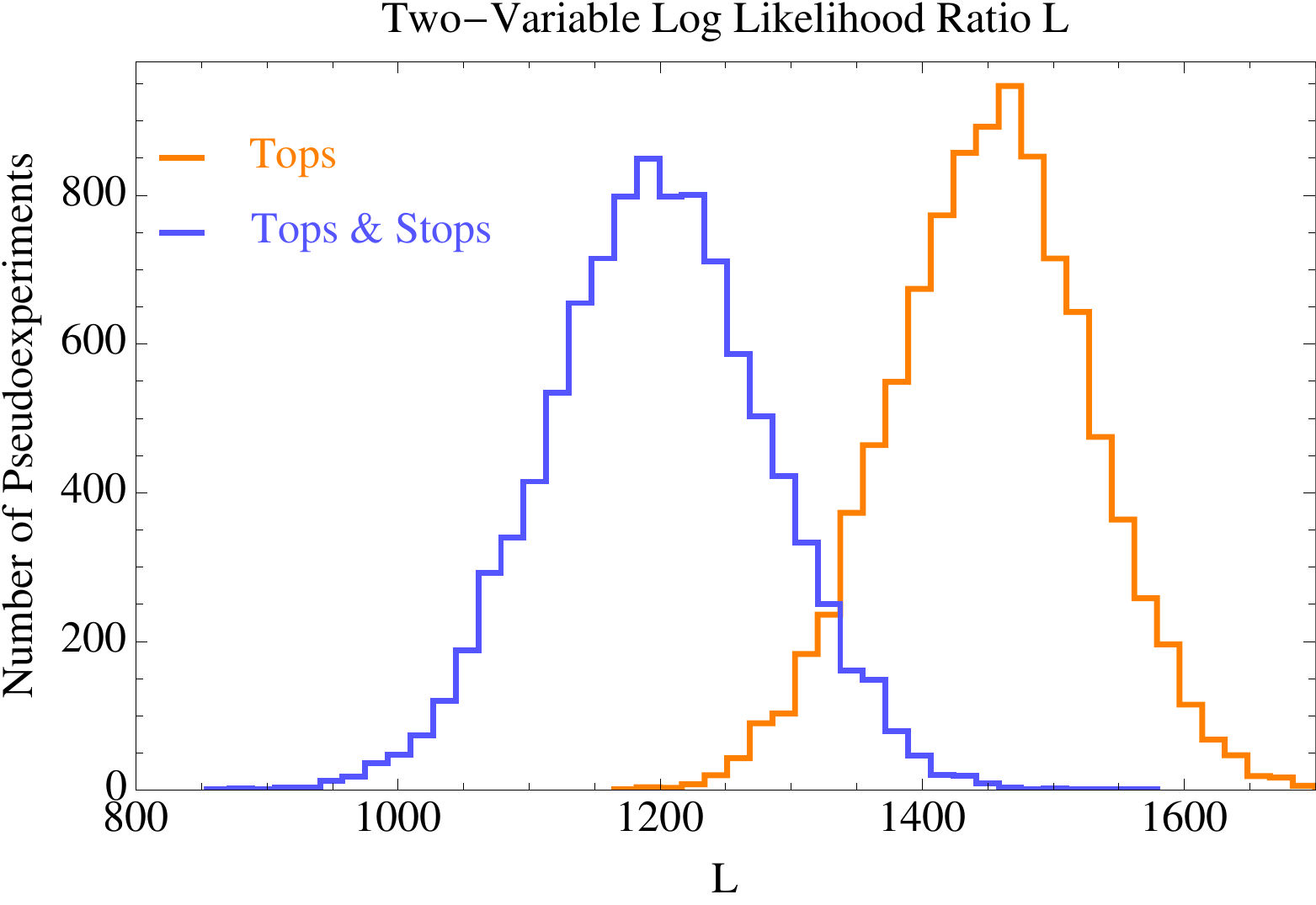}
\caption{The log likelihood ratio $L$ by combining two variables: $R$ and $\Delta y_{t\bar t}$. Each point on the curve corresponds to a pseudoexperiment with 32.8k events.}
\label{fig:L_combined}
}

%\section{Don't Stop the Beat: Boosted Stops}
\section{Boosted Stops}
\label{sec:boosted}
%%%%%%%%%%%%%%%%%%%%%%%%%%%%%%%%%%%%%%%%%%%%%%%%%%%%%%%%%

As discussed in \Sec{sec:current}, while measurements of the top production rate are in principle sensitive to stop production, 
difficulties arise because of challenging systematic uncertainties.  Indeed, even at NNLL theoretical uncertainties in the top cross section
are comparable to the size of the contributions we expect from stops.

\FIGURE{
\includegraphics[width=1\textwidth]{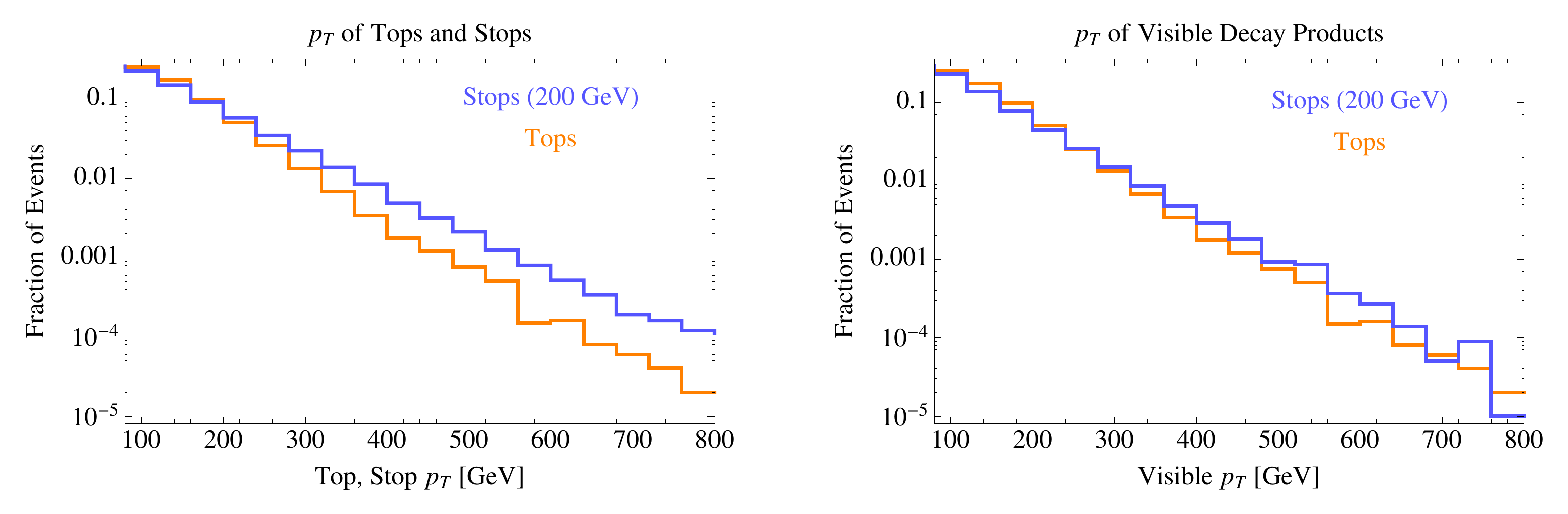}
\caption{Parton-level $p_T$ distributions of tops and stops (left) and of the corresponding visible decay products (right).  These distributions are normalized so that the top and stop share the same inclusive rate.  Note that here we have taken $m_{\tilde{t}}=200~{\rm GeV}$. }
\label{fig:ptcompare}
}

However, the physics which suppresses the stop production rate relative to that of tops  ({\it i.e.} phase space suppression from the larger stop mass, $p$-wave suppression, and the absence of 
$t$ and $u$-channel poles - see \Sec{sec:rapidity}) tends to have a diminished effect at higher $p_T$'s.  This  can be seen by comparing  normalized stop/top $p_T$ distributions, as shown  on the left hand side of \Fig{fig:ptcompare}.
One might therefore expect that through the use of top-tagging techniques (see \Ref{Plehn:2011tg} and references contained therein) one could observe a significant increase in top production at high $p_T$ coming from the decay of boosted stops.  

Alas, this is not the case for the stealth region we have considered.  The distribution of top/stop $p_T$'s is steeply falling,  and at high boosts the transverse momentum carried away by the LSP is enough to shift the distribution in $p_T$ of the 
visible decay products downward enough to largely cancel any hoped-for benefits (the fractional change in $p_T$ is roughly $\Delta m_{\tilde{t},t}/m_t$).  See the right-hand side of \Fig{fig:ptcompare}.  For this reason we do not include a more detailed study of boosted stops, although they can be useful outside of the stealthy regime.

%\bibliography{stopsmerged}
%\bibliographystyle{jhep}
\providecommand{\href}[2]{#2}\begingroup\raggedright\endgroup

\end{document}